\documentclass[12pt]{article}
\usepackage{amsmath,amssymb,amsthm}
\usepackage{bbm}
\begin{document}

\title{Eigenvalue degeneracy \\ in sparse random matrices}

\author{Masanari Shimura} 

\date{} 

\maketitle 

\begin{center} 
\it Graduate School of Mathematics, Nagoya University, Chikusa-ku,
\\  Nagoya 464-8602, Japan \\
\end{center}
\begin{center}
E-mail: shimura.masanari.s4@s.mail.nagoya-u.ac.jp
\end{center}

\begin{abstract}
In random matrices with independent and continuous matrix entries, the degeneracy 
probability of the eigenvalues is known to be zero. In this paper, random matrices including 
discontinuous matrix entries are analyzed in order to observe how degeneracy is generated. 
Using Erd{\"o}s-R{\'e}nyi matching probability theory of random bipartite graphs, 
we asymptotically evaluate the degeneracy probability of such random matrices.
As a  result, due to accumulation of the eigenvalues to the origin, a positive degeneracy 
probability is found for eigenvalues of a sparse random matrix model. 

\end{abstract}

\medskip

Keywords: \par
eigenvalue degeneracy; random matrices; sparse matrices;
random graphs.


\newpage

\section{Introduction}
\setcounter{equation}{0}
\renewcommand{\theequation}{1.\arabic{equation}}

In random matrix theory, eigenvalue distributions are often analyzed under the assumption that  
there is no eigenvalue degeneracy. This assumption enables diagonalizations of matrices with 
one-to-one correspondence of parameters,  and the implicit function theorem is applied to observe the eigenvalue motion in reaction to a matrix motion. Such an argument is possible because there 
is a common sense that eigenvalue degeneracy of random matrices does not take place in almost all situations. As explained in Tao's book\cite{TAO} on random matrices, this common sense is based on the following fact.
\par\bigskip\noindent
{\bf Proposition 1.1}
\par\smallskip\noindent 
Let us set $N \geq 2$. In the space of $N \times N$ hermitian matrices, the codimension of the 
space of matrices with at least one eigenvalue degenerated is $3$. In the space of $N \times N$ 
real symmetric matrices, the codimension of the space of matrices with at least one eigenvalue 
degenerated is $2$.
\par\bigskip\noindent
This proposition appeared in the paper\cite{JNW} by von Neumann and Wigner published in 1929. 
Let us explain it for a hermitian matrix. Suppose that an $N \times N$ hermitian matrix has $f$ 
distinct eigenvalues and the multiplicity of each distinct eigenvalue is $g_1,\cdots,g_f$. Then such a hermitian 
matrix can be written in terms of $N^2 + f - g_1^2 - g_2^2 - \cdots - g_f^2$ parameters. In a special case 
$g_1 = g_2 = \cdots = g_{f-1} = 1$ with $g_f=2$, we need $N^2 - 3$ parameters. As a  hermitian matrix in general is written in terms of $N^2$ parameters, the codimension is $3$.  Therefore, if the probability measure of the whole space of hermitian matrices is absolutely continuous with respect to the $N^2$ dimensional Lebesgue measure, the degeneracy probability of the matrix eigenvalues is zero.
\par\smallskip
On the other hand, when the probability measure is discontinuous, the degeneracy probability of random matrix eigenvalues has been examined relatively recently. A paper\cite{TV} by Tao and Vu in 2017 considered the case when matrix entries (upper-triangular and diagonal entries)  of a real symmetric matrix were independently distributed. The probability measure of each entry was not necessarily continuous. If the measure of a one point set was less than a constant independent of the matrix dimension 
$N$, the degeneracy probability of the eigenvalues converged to zero in the limit $N \rightarrow \infty$. 
Tao and Vu also studied  Erd\"os-R\'enyi random graphs with $N$ vertices.  When the probability to find 
a couple of vertices connected by an edge was independent of $N$, they showed that the eigenvalue 
degeneracy probability of the adjacency matrices approached to zero in the limit $N \rightarrow \infty$. 
Moreover, in the paper\cite{LV} by Luh and Vu in 2020,  sparse $N \times N$ real symmetric matrices were 
investigated. In the limit $N \rightarrow \infty$, they assumed that the probability to have a zero 
matrix entry converged to one, and found a sufficient condition to make the eigenvalue degeneracy 
probability tend to zero. 
\par\smallskip
In this paper, we analyze random matrix models with discontinuous probability measures in order to observe how 
degeneracy is generated. In particular, for a sparse random matrix model, we are able to find a positive 
degeneracy probability for the eigenvalues. Degeneracy occurs because of accumulation of the eigenvalues 
to the origin, and the accumulation comes from the discontinuity. 
\par\smallskip
The outline of this paper is the following. 
In \S 2,  we consider an $N \times N$ matrix which is an analytic function $\varphi(X_1,X_2,\cdots,X_N)$ of $d$ 
independent and continuous random variables $X_1,X_2,\cdots,X_d$. When there exists at least one family of 
constants $x_1,x_2,\cdots,x_d$ making the $N$ eigenvalues of  $\varphi(x_1,x_2,\cdots,x_d)$ all distinct, then 
the $N$ eigenvalues of $\varphi(X_1,X_2,\cdots,X_d)$ are all distinct with probability one. Thus we find that 
the eigenvalue degeneracy probability turns out to be either zero or one. This result can be applied to a variety 
of typical random matrix models including random unitary matrices. 
\par\smallskip
In \S 3, matrix models related to bipartite graphs are introduced. In these models, if the $j$-th and $\ell$-th vertices 
are not connected by an edge in the bipartite graph, then the $(j,\ell)$ entry of the corresponding 
matrix is fixed to zero. The concept of perfect matching is borrowed 
from the theory of bipartite graphs and used in the study of degeneracy. If a bipartite graph has a perfect matching, the eigenvalues of at least one 
of the corresponding matrices do not degenerate. Such an analysis using the concept of graph theory is also 
applied to symmetric matrices. 
Moreover, suppose that
$\phi$ is an $N \times N$ matrix with elements
\begin{equation}
  \phi_{j \ell} =
  \begin{cases}
    x_{j \ell}, & (j, \ell) \in \Lambda, \\
    0, & \text{otherwise}.
  \end{cases}
\end{equation}
Here $x_{j \ell} \in \mathbb{R} \text{ or } \mathbb{C}$
and $\Lambda \subset \{(j,\ell) \mid 1 \leq j,\ell \leq N\}$.
Then, if the cardinality of $\Lambda$ is larger than $N^2 - 2N + 1$,
a family of $x_{j\ell}$ ($(j,\ell) \in \Lambda$) exists,
so that the eigenvalues of $\phi$ do not degenerate.
\par\smallskip
In \S 4, employing the results of \S 2 and \S 3 together, we study the eigenvalue degeneracy probability for $N \times N$ sparse random matrices. Let us suppose that 
random variables $Y_{j \ell}$ ($j,\ell=1,2,\cdots,N)$ independently obey Bernoulli distribution 
with a parameter $p$. That is, $Y_{j \ell}$ takes value 
$1$ with probability $p$ and $0$ with probability $1 - p$. Here $p$ depends on $N$ as
\begin{equation}
p = p(N) = \frac{\log N + c}{N} + o\left( \frac{1}{N} \right), \ \ \ N \rightarrow \infty
\end{equation}
with a real $c$ fixed. Moreover we assume that continuous random variables $Z_{j \ell}$ ($j,\ell = 1,2,\cdots,N$) are 
independently distributed and consider the eigenvalue degeneracy probability of an 
$N \times N$ sparse random matrix $(Y_{j \ell} Z_{j \ell} )_{j,\ell=1,2,\cdots,N}$. The random 
matrix entries are continuously distributed except at the origin. In order to asymptotically 
evaluate the degeneracy probability in the limit $N \rightarrow \infty$, we refer to a theory by 
Erd\"os and R\'enyi concerning the perfect matching probability of random bipartite graphs.  
As a result, a positive asymptotic limit
\begin{equation}
1 - e^{-\lambda} - \lambda e^{-\lambda}
\end{equation}
is worked out ($\lambda = 2 e^{-c}$) for the degeneracy probability. Accumulation of the eigenvalues to the origin, due to the discontinuity of the matrix entry distribution, causes the degeneracy. In \S 5, we consider the degeneracy probability for symmetric sparse random matrices, 
and again find a positive asymptotic limit.

\section{Eigenvalue degeneracy probability for continuous distributions}
\setcounter{equation}{0}
\renewcommand{\theequation}{2.\arabic{equation}}

Matrix eigenvalues are roots of the characteristic equation.
In order to study the root degeneracy, 
we first introduce the concept of a discriminant.
The discriminant detects multiple roots and is expressed 
as the resultant of a polynomial and its derivative, 
which is given by the determinant of the Sylvester matrix 
(see, e.g., Gelfand-Kapranov-Zelevinsky~\cite{GKZ}, Ch.~12, §1).
Using discriminants, we show a sufficient condition for the eigenvalue degeneracy 
probability of random matrices to be zero or one.
\par
\medskip
\noindent
{\bf Definition 2.1}
\par
\medskip
\noindent
For an $n$-th order polynomial
\begin{equation}
p(\lambda) = \lambda^n + a_{n-1} \lambda^{n-1} + \cdots + a_1 \lambda + a_0,
\end{equation}
we set $b_k = k a_k$, $k=1,2,\cdots,n-1$. Then we define the 
discriminant $\Delta[p(\lambda)]$ of $p(\lambda)$ as the following determinant 
(multiplied with a coefficient) 
\begin{equation}
\Delta[p(\lambda)] = (-1)^{n(n-1)/2} \left| \begin{array}{ccccccccc}
1 & a_{n-1} & \cdots & \cdots & a_1 & a_0 & & & \\ 
 & 1 & a_{n-1} & \cdots & \cdots  & a_1 & a_0 &  & \\ 
 & & \ddots & \ddots & \ddots & \ddots & \ddots & \ddots &  \\ 
 &  &  & 1 & a_{n-1} & \cdots & \cdots & a_1 & a_0 \\ 
n & b_{n-1} & \cdots & b_2 & b_1 &  & & & \\ 
  & n & b_{n-1} & \cdots & b_2 & b_1 & & & \\ 
  & & \ddots &  \ddots & \ddots & \ddots & \ddots &  & \\ 
 & &  & n & b_{n-1} & \cdots & b_2 & b_1 & \\ 
 & &  & & n & b_{n-1} & \cdots & b_2 & b_1
\end{array} \right|
\end{equation}
of a $(2 n - 1) \times (2 n - 1)$ matrix. The upper $n-1$ rows consist  of the coefficients 
$1,a_{n-1},a_{n-2},\cdots,a_1,a_0$ of $p(\lambda)$  with one column shifted to the right each row. 
The lower $n$ rows consist of the coefficients $n,b_{n-1},b_{n-2},\cdots,b_2,b_1$ of $p'(\lambda)$ 
with one column shifted to the right each row. All vacancies are occupied with zeros. 
\par
\medskip
\noindent
{\bf Proposition 2.2}
\par
\medskip
\noindent
Let us suppose that the roots of an $n$-th order polynomial $p(\lambda) = \lambda^n + a_{n-1} \lambda^{n-1} 
+ \cdots + a_1 \lambda + a_0$ are $\lambda_1,\lambda_2,\cdots,\lambda_n$. Then the discriminant 
$\Delta[p(\lambda)]$ is equal to 
\begin{equation}
\prod_{1 \leq  j < \ell \leq n} (\lambda_j - \lambda_\ell)^2.
\end{equation}
\par
\medskip
\noindent
{\bf Proof}
\par
\medskip
\noindent
It follows from {\bf Definition 2.1} that the discriminant is a polynomial of the coefficients $a_0,a_1,\cdots,a_{n-1}$. The coefficients $a_0,a_1,\cdots,a_{n-1}$ are elementary symmetric polynomials of the roots $\lambda_1,\lambda_2,\cdots,\lambda_n$. Therefore the discriminant $\Delta[p(\lambda)]$ is a symmetric polynomial 
of the roots $\lambda_1,\lambda_2,\cdots,\lambda_n$. Let us set $\Delta[p(\lambda)] = (-1)^{n(n-1)/2} 
d(\lambda_1,\lambda_2,\cdots,\lambda_n)$.
\par
Now we consider an equality
\footnotesize
\begin{eqnarray}
& & \left( \begin{array}{cccccccc} 
1 & a_{n-1} & \cdots & \cdots & a_0 & & & \\ 
 & 1 & a_{n-1} & \cdots & \cdots  & a_0 &  & \\ 
 & & \ddots & \ddots & \ddots & \ddots & \ddots &  \\ 
 &  &  & 1 & a_{n-1} & \cdots & \cdots & a_0 \\ 
n & b_{n-1} & \cdots & b_2 & b_1 &  & & \\ 
  & n & b_{n-1} & \cdots & b_2 & b_1 & & \\ 
  & & \ddots &  \ddots & \ddots & \ddots & \ddots &  \\ 
 &  & & n & b_{n-1} & \cdots & b_2 & b_1
\end{array} \right)
\left( \begin{array}{cccccc}
1 & & & \lambda_1^{2 n - 2} & \cdots & \lambda_n^{2 n - 2} \\ 
  & \ddots & & \vdots  & \ddots & \vdots  \\
 & & 1 & \lambda_1^{n} & \cdots & \lambda_n^{n} \\
  & & & \lambda_1^{n-1} & \cdots & \lambda_n^{n-1} \\
 & & & \vdots & \ddots & \vdots \\
 & & & \lambda_1  & \cdots & \lambda_n \\ 
& & & 1 & \cdots & 1 \\ 
\end{array} \right) \nonumber \\ 
& = & \left( \begin{array}{cccccccc}
1 & * & \cdots & * & & & & \\
  & 1 & \ddots & \vdots & & & & \\ & & \ddots & * & & & & \\ &  & & 1 & & & & \\ 
  {}* & * & \cdots & * & 
  \lambda_1^{n-1} p'(\lambda_1) & \lambda_2^{n-1} p'(\lambda_2) & \cdots & \lambda_n^{n-1} p'(\lambda_n)  \\ 
  {}* & * & \cdots & * & 
  \lambda_1^{n-2} p'(\lambda_1) & \lambda_2^{n-2} p'(\lambda_2) & \cdots & \lambda_n^{n-2} p'(\lambda_n)  \\ 
  {}*  & * &  \ddots & * & \vdots & \vdots & \ddots & \vdots \\ 
  {}*  & * &  \cdots & * & p'(\lambda_1) & p'(\lambda_2) & \cdots & p'(\lambda_n)  \\
\end{array} \right)
\end{eqnarray}
\normalsize
and compute the determinant of the both sides, we find
\begin{equation}
d(\lambda_1,\lambda_2,\cdots,\lambda_n) 
\prod_{1 \leq  j < \ell \leq n} (\lambda_j - \lambda_\ell) = 
\prod_{j=1}^n p'(\lambda_j) \prod_{1 \leq  j < \ell \leq n} (\lambda_j - \lambda_\ell).
\end{equation}
Dividing the both sides with $\prod_{1 \leq  j < \ell \leq n} (\lambda_j - \lambda_\ell)$ yields
\begin{eqnarray}
d(\lambda_1,\lambda_2,\cdots,\lambda_n) & = & 
\prod_{j=1}^n p'(\lambda_j) = \prod_{j=1}^n \prod_{\ell (\neq j)} (\lambda_j - \lambda_\ell) 
\nonumber \\ 
& = & 
(-1)^{n(n-1)/2} \prod_{1 \leq  j < \ell \leq n} (\lambda_j - \lambda_\ell)^2,
\end{eqnarray}
which gives the required result. \qed
\par
Because of {\bf Proposition 2.2}, a  polynomial $p(\lambda)$ has a multiple root, if and only if the discriminant 
$\Delta[p(\lambda)]$ is zero.
\par
\medskip
\noindent
{\bf Definition 2.3}
\par
\medskip
\noindent
Suppose that $K = \mathbb{R}$ or $K = \mathbb{C}$. Let us call a function $f$ from an open set in $K^d$ to 
$\mathbb{C}$ analytic, if there is a Taylor expansion in the form
\begin{equation}
f(x) = \sum_{k=0}^\infty \sum_{|\alpha| = k} c_\alpha (x - a)^\alpha
\end{equation}
with $x = (x_1,x_2,\cdots,x_d)$ in the neighborhood of each $a \in K^d$ within the domain. 
Here $\alpha = (\alpha_1,\alpha_2.\cdots,\alpha_d)$ moves within 
$\mathbb{Z}_+^d := \{ \alpha \in \mathbb{Z}^d: \alpha_j \geq 0, \ j = 1,2,\cdots,d \}$, 
$c_{\alpha} \in \mathbb{C}$ and
\begin{equation}
|\alpha| = \alpha_1 + \alpha_2 + \cdots \alpha_d, \ \ \ x^\alpha= x_1^{\alpha_1} x_2^{\alpha_2} \cdots x_d^{\alpha_d}.
\end{equation}
Moreover we call a function
\begin{equation}
F(x) = (f_1(x),f_2,(x),\cdots,f_n(x))
\end{equation}
from an open set in $K^d$ to $\mathbb{C}^n$ analytic, if $f_1,f_2,\cdots,f_n$ are all analytic.
\par
\medskip
\noindent
Supposing that $x_1,x_2.\cdots,x_{d-1}$ are constants, we obtain an analytic 
function $f_1(x) = f(x_1,x_2,\cdots,x_{d-1},x)$ of one variable $x \in K$. As is well-known, 
when $K = \mathbb{C}$, the zero set of an analytic function of one variable 
is an at most countable set or the whole domain. In the case $K = \mathbb{R}$, 
the zero set is also an at most countable set or the whole domain, because 
each of the real and imaginary parts of an analytic function with $K = \mathbb{R}$ 
can be analytically cotinuated to the complex plane.
\par
Let us now introduce a probability space $(\Omega, {\cal F},\mathbb{P})$ with a sample space 
$\Omega$, an event space ${\cal F}$ and a  probability function $\mathbb{P}$. 
\par
\medskip
\noindent
{\bf Definition 2.4}
\par
\medskip
\noindent
Suppose again that $K = \mathbb{R}$ or $K = \mathbb{C}$. If a $K$-valued random variable $X$ satisfies
\begin{equation}
\forall x \in K, \ \ \ \mathbb{P}(X=x) = 0,
\end{equation}
then $X$ is called a broad continuous random variable (or simply a continuous random variable).
\par
In particular, a random variable with an absolutely continuous measure is broad continuous. 
A broad continuous random variable is not necessarily absolutely continuous. 
\par
In the following, the set of all $N \times N$ matrix on $\mathbb{C}$ is denoted by $\mathbb{M}_N(\mathbb{C})$.
\par
\medskip
\noindent
{\bf Theorem 2.5}
\par
\medskip
\noindent
Let us put $K = \mathbb{R} $ or $K = \mathbb{C}$. Suppose that $D_1,D_2,\cdots,D_d$ are 
open sets on $K$ and $D = D_1 \times D_2 \times \cdots \times D_d$. We introduce 
independent and (broad) continuous random variables $X_1,X_2,\cdots,X_d$: $\Omega \rightarrow K$, and assume that $\mathbb{P}(X_j \in  D_j) = 1$ for each $j=1,2,\cdots,d$. Let us consider a random 
matrix $\varphi(X_1,X_2,\cdots, X_d)$ given by an analytic function $\varphi(x_1,x_2,\cdots,x_d)$ ($\varphi: D \rightarrow 
\mathbb{M}_N(\mathbb{C})$). If at least one vector $(x_1,x_2,\cdots,x_d) \in D$ exists and $\varphi(x_1,x_2,\cdots,x_d)$ has $N$ distinct 
eigenvalues, then the random matrix $\varphi(X_1,X_2,\cdots,X_d)$ has $N$ distinct eigenvalues 
with probability one.
\par
\medskip
\noindent
{\bf Remark}
\par
\medskip
\noindent
If there exits at least one pair of multiple eigenvalues of $\varphi(x_1,x_2,\cdots,x_d)$ for arbitrary $(x_1,x_2,\cdots,x_d) \in D$, then the probability for $\varphi(X_1,X_2,\cdots,X_d)$ to have $N$ distinct eigenvalues 
is obviously zero. That is, the probability to have $N$ distinct eigenvalues is in general either zero or one. 
\par
\medskip
\noindent
{\bf Proof of Theorem 2.5}
\par
\medskip
\noindent
We denote the distribution measure of $X_j$ by $\mu_j$ for $j=1,2,\cdots,d$. Since $X_j$  is (broad)  continuous, for an at most countable set $A \subset D_j$, $\mu_j(A) = 0$. Let us write the discriminant of the characteristic polynomial of a matrix $P$ as $\Delta(P)$. Because $\Delta(P)$ is a polynomial whose indeterminates are the entries of the matrix $P$, the function $\Delta(\varphi(x_1,x_2,\cdots,x_d))$ is analytic. As $\Delta(\varphi(x_1,x_2,\cdots,x_d))$ is analytic, when $x_1 \in D_1$, $x_2 \in D_2$,$\cdots$, $x_{d-1} \in D_{d-1}$ are arbitrarily fixed, the zero set
\begin{equation}
\{ x \in D_d | \Delta(\varphi(x_1,x_2,\cdots,x_{d-1},x)) = 0 \}
\end{equation}
is an utmost countable set or the whole $D_d$.  Therefore we find
\begin{eqnarray}
& & \int_{D_d} \ \mathbbm{1}_{\Delta (\varphi(x_1,\cdots,x_{d-1},x)) \neq 0} \  d\mu_d(x) 
\nonumber \\ & = & 
\left\{ \begin{array}{ll} 1, & \exists x_d, \ \Delta(\varphi(x_1,\cdots,x_{d-1},x_d)) \neq 0, \\ 
0, & {\rm otherwise}. \end{array} \right.
\end{eqnarray}
Repeatedly using this relation leads to
\begin{eqnarray}
& & \mathbb{P}\left(\varphi(X_1,X_2,\cdots,X_d) \ {\rm has} \ N \ {\rm distinct} \ {\rm eigenvalues} \right) \nonumber \\ 
& & = \int_{D_1} d\mu_1 \int_{D_2} d\mu_2 \cdots \int_{D_d} d\mu_d \ 
\mathbbm{1}_{\Delta(\varphi) \neq 0}  \nonumber \\ 
& & = 
\left\{ \begin{array}{ll} 1, & \exists \{x_1,x_2,\cdots,x_d \}, \ \Delta(\varphi(x_1,x_2,\cdots,x_d)) \neq 0, \\ 
0, & {\rm otherwise}, \end{array} \right.
\end{eqnarray}
which gives the required result. \qed
\par
\medskip
\noindent
{\bf Example 2.6}
\par
\medskip
\noindent
Let us put $\mathbbm{i} = \sqrt{-1}$. The whole set of $N \times N$ hermitian matrices is identified 
with $\mathbb{R}^{N^2}$ by means of the following map $\varphi: \mathbb{R}^{N^2} \rightarrow \mathbb{M}_N(\mathbb{C})$.
\begin{eqnarray}
& & \phi(x_{11},\cdots,x_{NN}) \nonumber \\ 
& = & \left( \begin{array}{cccc} x_{11} & x_{12} & \cdots & x_{1N} \\ 
x_{12} & x_{22} & \cdots & x_{2N} \\ 
\vdots & \vdots & \ddots & \vdots \\ 
x_{1N} & x_{2N} & \cdots & x_{NN} 
 \end{array} \right)
+ \mathbbm{i} 
\left( \begin{array}{ccccc} 0 & -x_{21} & \cdots & & -x_{N1} \\ 
x_{21} & 0 & \cdots & & -x_{N2} \\ 
\vdots & \vdots & \ddots & & \vdots \\ 
x_{N1} & x_{N2} & \cdots & & 0 
 \end{array} \right). \nonumber \\
 \end{eqnarray}
We define the subset $S \subset \mathbb{R}^{N^2}$ as the whole of the $N \times N$ hermitian matrices with multiple eigenvalues. Then one can prove that the Lebesgue measure 
$\mathfrak{m}(S)$ is zero\cite{MOF}. This fact can also be proved by using {\bf Theorem 2.5}, as explained below. 
\par
Let us arbitrarily fix $n \in \mathbb{N}$ and set $I_n = [-n,n]$. If each of independent random 
variables $X_{11},\cdots,X_{NN}$ is distributed according to a probability density function
\begin{equation}
p(x) = \left\{ \begin{array}{ll} 1/(2 n), & x \in I_n, \\ 0, & x \notin I_n, \end{array} \right.
\end{equation}
the probability for $\varphi(X_{11},\cdots,X_{NN})$  to have multiple eigenvalues is equal to 
$(2 n)^{-N^2} \mathfrak{m}(S \cap I_n^{N^2})$. On the other hand, if we put $x_{j \ell} = j \delta_{j \ell}$ 
($\delta_{j \ell}$ is the Kronecker delta), the eigenvalues of $\varphi(x_{11}, \cdots,x_{NN})$ 
do not degenerate. Then it follows from {\bf Theorem 2.5} that the probability for the eigenvalues 
of $\varphi(X_{11},\cdots,X_{NN})$ to degenerate is zero. Therefore $\mathfrak{m}(S \cap I_n^{N^2})= 0$ holds for 
arbitrary $n \in \mathbb{N}$ and we obtain
\begin{equation}
\mathfrak{m}(S) = \mathfrak{m}\left( \bigcup_{n=1}^\infty \left(S \cap I_n^{N^2} \right) \right) = \lim_{n \rightarrow \infty} 
\mathfrak{m}\left(S \cap I_n^{N^2} \right) = 0.
\end{equation}
\par
\medskip
\noindent
{\bf Example 2.7}
\par
\medskip
\noindent
Suppose that $E_{j \ell} \in \mathbb{M}_N(\mathbb{C})$ is a matrix unit, so that 
the $(j,\ell)$ entry of $E_{j \ell}$ is one and the other entries are all 
zero. We define
\begin{equation}
P_j = E_{jj}, \ \ \ Q_{j \ell} = - \mathbbm{i} E_{j \ell} + \mathbbm{i} E_{\ell j}
\end{equation}
and an analytic function $\varphi: \mathbb{R}^{N^2} \rightarrow \mathbb{M}_N(\mathbb{C})$ as 
\begin{eqnarray}
& & \varphi(x_{11},\cdots,x_{NN}) \nonumber \\ 
& = & \left( \prod_{j=1}^{N-1} \prod_{\ell = j + 1}^N {\rm exp}(\mathbbm{i} x_{\ell j} P_{\ell} ) \ 
{\rm exp}(\mathbbm{i} x_{j \ell} Q_{j \ell}) \right) \ \left(\prod_{j=1}^N {\rm exp}(\mathbbm{i} x_{jj} P_j) \right),
\end{eqnarray}
where the order of a product of matrices is defined as $\prod_{j=1}^n A_j = A_1 A_2 \cdots A_n$. If
the probability density function $p_{11},\cdots,p_{NN}$ for each of independent $\mathbb{R}$-valued random variables $X_{11},\cdots,X_{NN}$ is
\begin{align}
p_{j \ell}(x)  = & 1/(2 \pi),  & x \in [0, 2 \pi] \ {\rm for} \ j \geq \ell, \nonumber \\ 
p_{j \ell}(x) =  & 2 (\ell - j) \ \sin x \ \cos^{2 (\ell - j)-1}x, & x \in [0,\pi/2] \ {\rm for} 
\ j < \ell,
\end{align}
then the probability measure of $\varphi(X_{11},\cdots,X_{NN})$ is the Haar measure 
of the unitary group\cite{SHH}. In a special case $x_{j \ell} = 2 \pi j \delta_{j \ell}/N$, the eigenvalues 
of $\varphi(x_{11},\cdots,x_{NN})$ do not degenerate. Therefore we can see from {\bf Theorem 2.5} that the probability for the eigenvalues of $\varphi(X_{11},\cdots,X_{NN})$ to degenerate is zero.
\par
\medskip
\noindent
{\bf Lemma 2.8}
\par
\medskip
\noindent
We set $K = \mathbb{R}$ or $K = \mathbb{C}$. Let us consider a $K$-linear map $\varphi: K^d \rightarrow \mathbb{M}_N(\mathbb{C})$. Suppose that a set of constants $c_1,\cdots,c_r \in K$ exists for $r < d$, so that $\varphi(c_1,\cdots,c_r,x_{r+1},\cdots,x_d)$ has at least one pair of multiple 
eigenvalues for arbitrary $x_{r+1},\cdots,x_d$.  Then $\varphi(0,\cdots,0,x_{r+1},\cdots,x_d)$ 
(the constant arguments $c_1,\cdots,c_r$ of $\varphi$ are replaced with $0,\cdots,0$) has at least one pair of multiple eigenvalues. 
\par
\medskip
\noindent
{\bf Proof}
\par
\medskip
\noindent
We introduce notations
\begin{equation}
y = (y_1,\cdots,y_r), \ \ \ {\tilde x} = (x_{r+1},\cdots,x_d)
\end{equation}
and denote the discriminant of the characteristic polynomial of a matrix $P$ by $\Delta(P)$.
Because $\varphi(x_1,\cdots,x_d)$ is a linear homogeneous polynomial of $x_1,\cdots,x_d$ 
and $\Delta(P)$ is a degree $N(N-1)$ homogeneous polynomial of the entries of 
$P$, we can write $\Delta(\varphi(y,{\tilde x}))$ in the form
\begin{equation}
\Delta(\varphi(y,{\tilde x}))= \sum_{|\alpha| \leq N(N-1)} \sum_{|\beta|=N(N-1)-|\alpha|} 
\kappa_{\beta \alpha} y^{\beta} {\tilde x}^{\alpha}, \ \ \ \kappa_{\beta \alpha} \in \mathbb{C}.
\end{equation}
The multiple index notation was introduced in {\bf Definition 2.3}. When $y=c=(c_1,\cdots,c_r)$, we 
assume that $\Delta(\varphi(c,{\tilde x}))= 0$ for arbitrary ${\tilde x}$. Therefore
\begin{equation}
\sum_{|\beta|=N(N-1)-|\alpha|} \kappa_{\beta \alpha} c^{\beta} = 0
\end{equation}
for arbitrary $\alpha$ with $|\alpha| \leq N(N-1)$.  In a special case  $|\alpha|=N(N-1)$, we obtain $\kappa_{(0,\cdots,0) \alpha} = 0$. Then it follows that
\begin{eqnarray}
\Delta(\varphi(0,\cdots,0,{\tilde x})) & = & 
\sum_{|\alpha| = N(N-1)} 
\kappa_{(0,\cdots,0) \alpha}  {\tilde x}^{\alpha} \nonumber \\ 
& & + 
\sum_{|\alpha| < N(N-1)} \sum_{|\beta|=N(N-1)-|\alpha|} 
\kappa_{\beta \alpha} (0,\cdots,0)^{\beta} {\tilde x}^{\alpha} \nonumber \\ 
& = & 0,
\end{eqnarray}
which gives the required result. \qed
\par
Using {\bf Lemma 2.8}, we are able to prove the following {\bf Corollary 2.9}, which generalizes 
{\bf Theorem 2.5} for a  linear map $\varphi$ on $\mathbb{R}^d$. It gives a sufficient condition for a random matrix $\varphi(X_1,\cdots,X_d)$ 
to have $N$ distinct eigenvalues, when some of $X_1,\cdots,X_d$ are not necessarily continuous.
\par
\medskip
\noindent
{\bf Corollary 2.9}
\par
\medskip
\noindent
Let $K=\mathbb{R}$ or $K=\mathbb{C}$. Suppose that $\varphi: K^d \rightarrow \mathbb{M}_N(\mathbb{C})$ is a $K$-linear map. 
Moreover we assume that a vector $(x_{r + 1},\cdots,x_d ) \in K^{d-r}$ exists for 
$r < d$, so that 
$\varphi(0,\cdots,0,x_{r+1}, \cdots, x_d)$ has $N$ distinct eigenvalues. If $X_1,\cdots,X_d:
\Omega \rightarrow K$ are independent random variables and $X_{r+1},\cdots,X_d$ 
are (broad) continuous, then $\varphi(X_1,\cdots,X_d)$ have $N$ distinct eigenvalues with 
probability one. 
\par
\medskip
\noindent
{\bf Proof}
\par
\medskip
\noindent
The contraposition of {\bf Lemma 2.8} implies that 
$\varphi(c_1,\cdots,c_r,x_{r+1},\cdots,x_d)$ has $N$ distinct eigenvalues for 
arbitrary $c_1,\cdots,c_r \in K$. Then {\bf Theorem 2.5} gives the 
required result, because $X_{r+1},\cdots,X_d$ are (broad) continuous.

\section{Eigenvalue degeneracy and perfect matching of graphs}
\setcounter{equation}{0}
\renewcommand{\theequation}{3.\arabic{equation}}

\subsection{Asymmetric matrices}

Let us first consider asymmetric matrices and corresponding bipartite graphs. 
In particular, we focus on the relation between eigenvalue degeneracy and 
perfect matching of graphs. In the following, $E_{j \ell}$ stands for a matrix 
unit defined in {\bf Example 2.7}.
\par
\medskip
\noindent
{\bf Lemma 3.1}
\par
\medskip
\noindent
Let us suppose that $\sigma$ is an arbitrary element of the $N$-th order symmetric group 
$\mathfrak{S}_N$ and define an $N \times N$ matrix
\begin{equation}
\varphi(x_1,x_2,\cdots,x_N) = \sum_{j=1}^N x_j E_{j \ \sigma(j)}, \ \ \ x_j \in K,
\end{equation}
where $K$ is $\mathbb{R}$ or $\mathbb{C}$. That is, the $(m,n)$ entry 
$\varphi_{mn}$ of $\varphi(x_1,x_2,\cdots,x_N)$ is 
\begin{equation}
\varphi_{mn} = \sum_{j=1}^N x_j \ \delta_{jm} \ \delta_{\sigma(j) n} = x_m \delta_{\sigma(m) n}.
\end{equation}
Then a set $\{ x_1,x_2,\cdots,x_N \}$ exists, for which the $N$ eigenvalues of $\varphi$  are 
different from each other and all non-zero. 
\par
\medskip
\noindent
{\bf Proof}
\par
\medskip
\noindent
A permutation $\sigma \in \mathfrak{S}_N$ can be written as a product of cyclic permutations involving 
no common indices. That is, denoting a cyclic permutation
$$
n_1 \rightarrow n_2 \rightarrow \cdots \rightarrow n_\ell \rightarrow n_1
$$
as $(n_1 n_2 \cdots n_\ell)$, we can write $\sigma$ in the form
\begin{equation}
\sigma = (n_{11} n_{12} \cdots n_{1 \ell_1}) (n_{21},n_{22},\cdots,n_{2 \ell_2}) \cdots 
(n_{r1} n_{r2} \cdots n_{r \ell_r}),
\end{equation}
where $1,2,\cdots,N$ is reordered to
\begin{equation}
n_{11}, n_{12},\cdots,n_{1 \ell_1},n_{21},n_{22},\cdots,n_{2 \ell_2}.\cdots,n_{r1},n_{r2},\cdots,n_{r \ell_r}.
\end{equation}
Now the characteristic polynomial of $\varphi$, which is the determinant of $\lambda I_N - \varphi$ 
with the $N \times N$ identity matrix $I_N$, can be factored as
\begin{equation}
\prod_{j=1}^r \left( \lambda^{\ell_j} - x_{n_{j 1}} x_{n_{j 2}} \cdots x_{n_{j \ell_j}} \right).
\end{equation}
Here each factor corresponds to a cyclic permutation $(n_{j1} n_{j2} \cdots n_{j \ell_j})$.
If we choose $x_1,x_2,\cdots,x_N$ such that 
$$
\sqrt[\ell_j]{ \left| x_{n_{j 1}} x_{n_{j 2}} \cdots x_{n_{j \ell_j}} \right|}, \ \ \ j = 1,2,\cdots r
$$
are different from each order and all non-zero, we obtain the required result. \qed
\par
\medskip
Now we introduce a terminology of graph theory. 
\par
\medskip
\noindent
{\bf Definition 3.2}
\par
\medskip
\noindent
A bipartite graph is a triple $({\cal V}_1,{\cal V}_2,{\cal E})$, where ${\cal V}_1$ and ${\cal V}_2$ 
are called vertex sets, and ${\cal E} \subset {\cal V}_1 \times {\cal V}_2$ is called an edge set. 
As for a bipartite graph $G = ({\cal V}_1,{\cal V}_2,{\cal E})$, if a subset $M$ of ${\cal E}$ satisfies
\begin{equation}
(v_1,v_2), (v'_1,v'_2) \in M \ {\rm and} \ (v_1,v_2) \neq (v'_1,v'_2) \Longrightarrow 
v_1 \neq v'_1 \ {\rm and} \ v_2 \neq v'_2,
\end{equation}
then $M$ is a matching of $G$. Moreover, if a matching $M$ meets a condition that 
$v_2 \in {\cal V}_2$ satisfying $(v_1,v_2) \in M$ exists for an arbitrary $v_1 \in {\cal V}_1$, then 
$M$ is a perfect matching of $G$. In that case $G$ is said to have a perfect matching $M$.
\par
In the following, we consider a bipartite graph $G = ({\cal V}^R, {\cal V}^C, {\cal E})$, supposing that
\begin{equation}
{\cal V}^R = \{ v_1^R,v_2^R,\cdots,v_N^R \}, \ \ \ {\cal V}^C = \{ v_1^C,v_2^C,\cdots,v_N^C \}, \ \ \ 
{\cal E} \subset {\cal V}^R \times {\cal V}^C.
\end{equation}
Let us  put ${\cal I}_N = \{1,2,\cdots N\}$. A subset $\Lambda$ of ${\cal I}_N \times {\cal I}_N $ is identified with 
a subset ${\cal E}$ of ${\cal V}^R \times {\cal V}^C$ by using a bijection
\begin{equation}
\Lambda \subset {\cal I}_N \times {\cal I}_N \longleftrightarrow {\cal E} = \{ (v_j^R,v_\ell^C)  | (j,\ell) \in \Lambda \} 
\subset {\cal V}^R \times {\cal V}^C.
\end{equation}
Hereafter we abbreviate $(v_j^R,v_\ell^C) \in {\cal E}$ as $\langle j,\ell \rangle$. We moreover write 
a family of parameters $x_{j \ell}$ with $(j, \ell) \in \Lambda$ as $x_\Lambda$.
\par
\medskip
\noindent
{\bf Lemma 3.3}
\par
\medskip
\noindent
Let us suppose that $\Lambda \subset {\cal I}_N \times {\cal I}_N$ and $\varphi(x_\Lambda) = 
\sum_{(j,\ell) \in \Lambda} x_{j \ell} E_{j \ell}$ with $x_{j \ell} \in K$ 
($K$ is $\mathbb{R}$ or $\mathbb{C}$). If a bipartite graph $G = ({\cal V}^R,{\cal V}^C, {\cal E})$ has a perfect matching, then an $x_\Lambda$ exits, so that all the $N$ eigenvalues of $\varphi(x_\Lambda)$ are different from each other and all non-zero.
\par
\medskip
\noindent
{\bf Proof}
\par
\medskip
\noindent
Let ${\cal M}$ be a perfect matching of $G$. From the definition of a perfect matching, for an arbitrary 
$v_j^R \in {\cal V}^R$, there exits a unique $v_\ell^C \in {\cal V}^C$ satisfying 
$\langle j,\ell \rangle \in {\cal M}$. We denote this $\ell$ by $\sigma(j)$.  Since 
$\sigma: j \mapsto \sigma(j)$ is an injection, $\sigma \in \mathfrak{S}_N$ and ${\cal M}$ can be written as
\begin{equation}
{\cal M}=  \{ \langle 1,\sigma(1) \rangle, \langle 2,\sigma(2) \rangle, \cdots, \langle N,\sigma(N) \rangle \}.
\end{equation}
Correspondingly, we also define
\begin{equation}
M =  \{ (1,\sigma(1)), (2,\sigma(2)), \cdots, (N,\sigma(N)) \}.
\end{equation}
Then, supposing that $x_{j \ell} = 0$ for $( j,\ell) \in \Lambda \setminus M$, we apply {\bf Lemma 3.1} to
\begin{equation}
{\tilde \varphi}(x_M) = \varphi(x_{\Lambda \setminus M} = 0, x_M) = \sum_{j=1}^N x_{j \sigma(j)} E_{j \sigma(j)}
\end{equation}
and obtain the required result. \qed
\par
\medskip
Corresponding to a bipartite graph $G = ({\cal V}^R,{\cal V}^C,{\cal E})$ and $S \subset {\cal I}_N$, 
we define a bipartite subgraph $G[S] = \left( {\cal V}^R[S],{\cal V}^C[S],{\cal E}[S] \right)$ as
\begin{equation}
{\cal V}^R[S] = \{ v_j^R | j \in S \}, \ \ \ 
{\cal V}^C[S] = \{ v_j^C | j \in S \}, \ \ \ 
{\cal E}[S] = \{ \langle j,\ell \rangle \in {\cal E} | j,\ell \in S \}.
\end{equation}
\par
\medskip
\noindent
{\bf Theorem 3.4}
\par
\medskip
\noindent
Let us suppose that $\Lambda \subset {\cal I}_N \times {\cal I}_N$ and  
$\varphi(x_\Lambda) = \sum_{(j,\ell) \in \Lambda} x_{j \ell} E_{j \ell}$ with $x_{j \ell} \in K$. Here $K$ 
is $\mathbb{R}$ or $\mathbb{C}$.
Then the following conditions are equivalent.
\par
\medskip
\noindent
(1) An $x_\Lambda$ exists, so that the $N$ eigenvalues of $\varphi(x_\Lambda)$ are different from each other. 
\par
\medskip
\noindent
(2) An $x_\Lambda$ exists, so that $\varphi(x_\Lambda)$ has no zero eigenvalues or a simple zero eigenvalue. Here a simple zero eigenvalue means an eigenvalue with multiplicity one.
\par
\medskip
\noindent
(3) The bipartite graph $G = ({\cal V}^R,{\cal V}^C,{\cal E})$ has a perfect matching, or 
$j \in {\cal I}_N$ exists and the bipartite subgraph $G[{\cal I}_N \setminus \{j\}]$ has 
a perfect matching.
\par
\medskip
\noindent
{\bf Proof}
\par
\medskip
\noindent
(1) $\Rightarrow$ (2): It is obvious.
\par
\medskip
\noindent
(2) $\Rightarrow$ (3): If $\varphi(x_\Lambda)$ has no zero eigenvalues, then
\begin{equation}
\det \varphi(x_\Lambda) = \det\left( \sum_{(j,\ell) \in \Lambda} x_{j \ell} E_{j \ell} \right) = 
\sum_{\sigma \in \mathfrak{S}_N} {\rm sign}(\sigma) \prod_{n=1}^N \left( 
\sum_{(j,\ell) \in \Lambda} x_{j \ell}  \delta_{j n} \delta_{\ell \sigma(n)} \right) \neq 0.
\end{equation}
Therefore there exists at least one $\sigma \in \mathfrak{S}_N$ satisfying $(n,\sigma(n)) \in \Lambda$ 
for all $n=1,2,\cdots,N$. This means that $G$ has a perfect matching. 
\par
Let us next consider the characteristic polynomial  $\varpi(\lambda) = 
\det( \lambda I_N- \varphi(x_\Lambda)$. If $\varphi(x_\Lambda)$ 
has a simple zero eigenvalue, $\varpi'(0) \neq 0$. Then at least one of 
the order $N-1$ principal minors of $\varphi(x_\Lambda)$ is non-zero, since the 
sum of those principal minors is $\varpi'(0)$. Then it follows that 
$j \in {\cal I}_N$ exists and $G[{\cal I}_N \setminus \{j\}]$ has a perfect matching.
\par
\medskip
\noindent
(3) $\Rightarrow$ (1): When $G$ has a perfect matching, it is obvious because of 
{\bf Lemma 3.3}. When $G[{\cal I}_N \setminus \{j\}]$ has a perfect matching for $j \in {\cal I}_N$, 
let us consider the $(N-1) \times (N-1)$ submatrix generated from $\varphi(x_\Lambda)$ by 
removing the $j$-th row and $j$-th column. We can see from {\bf Lemma 3.3} that there 
exists a family of parameters $x_{k \ell}$ with $\langle k,\ell \rangle 
\in {\cal E}[{\cal I}_N \setminus \{ j \}]$, 
so that the $N-1$ submatrix eigenvalues are 
different from each other and all non-zero. Moreover we put $x_{k \ell} = 0$ when 
$\langle k, \ell \rangle \in {\cal E}  \setminus {\cal E}[{\cal I}_N \setminus \{ j \}]$. Then 
the $N$ eigenvalues of $\varphi(x_\Lambda)$ are different from each other, because they consist 
of a simple zero eigenvalue and the submatrix eigenvalues. \qed
\par
\medskip
As an application, let us prove the following theorem.
\par
\medskip
\noindent
{\bf Theorem 3.5}
\par
\medskip
\noindent
Suppose that $N \geq 2$, $\Lambda \subset {\cal I}_N \times {\cal I}_N$ and $\varphi(x_\Lambda) = \sum_{(j,\ell) \in \Lambda} x_{j \ell} E_{j \ell}$ with $x_{jl} \in K$. Here $K$ is $\mathbb{R}$ or $\mathbb{C}$. If $\# \Lambda \geq 
N^2 - 2 N + 2$, there exits an  $x_\Lambda$, so that the $N$ eigenvalues of $\varphi(x_\Lambda)$ are different from each other. Here the number of the elements of a set $A$ is denoted by $\# A$.
\par
\medskip
\noindent
If $\# \Lambda < N^2 - 2 N + 2$, there are counterexamples. For example, if we have
\begin{equation}
\Lambda = \{ (j,\ell) | j \geq 3 \} \cup \{(2,1) \},
\end{equation}
then $\# \Lambda = N^2 - 2 N + 1$ and the zero eigenvalues of $\varphi(x_\Lambda)$ degenerate.
\par
In order to prove {\bf Theorem 3.5}, we use Hall's marriage theorem in graph theory. For a 
bipartite graph $G = ({\cal V}_1,{\cal V}_2,{\cal E})$, a subset $J_1 \subset {\cal V}_1$ and a 
subset $J_2 \subset {\cal V}_2$, we define
\begin{eqnarray}
\Gamma_G(J_1) & = & \{ v_2 \in {\cal V}_2 | \exists v_1 \in J_1, (v_1,v_2) \in {\cal E} \}, \nonumber \\ 
\Gamma_G(J_2) & = & \{ v_1 \in {\cal V}_1 | \exists v_2 \in J_2, (v_1,v_2) \in {\cal E} \}.
\end{eqnarray}
If $J_1$ or $J_2$ is a one-point set $\{ v \}$, then $\Gamma_G(\{ v \})$ is simply written as $\Gamma_G(v)$.
\par
\medskip
\noindent
{\bf Proposition 3.6} (Hall's marriage theorem)
\par
\medskip
\noindent
Let ${\cal V}_1$ be a finite set. For a bipartite graph $G = ({\cal V}_1,{\cal V}_2,{\cal E})$, the following conditions are equivalent.
\par
\medskip
\noindent
(a) A perfect matching exists.
\par
\medskip
\noindent
(b) For an arbitrary $A \subset {\cal V}_1$, $\# A \leq \# \Gamma_G(A)$.
\par
\medskip
\noindent
This (b) is called Hall's condition. This theorem is named after Hall who first published it in 1935\cite{HALL} and 
a proof of it can be found in \cite{LP} (Theorem 1.1.3). It leads to the following lemma.
\par
\medskip
\noindent
{\bf Lemma 3.7}
\par
\medskip
\noindent
Suppose that $G = ({\cal V}_1,{\cal V}_2,{\cal E})$ is a bipartite graph  with $\# {\cal V}_1 = \# {\cal V}_2 = n$. If $\# {\cal E} \geq n^2 - n + 1$, 
$G$ has a perfect matching.
\par
\medskip
\noindent
{\bf Proof}
\par
\medskip
\noindent
Let us arbitrarily take $A \subset {\cal V}_1$ ($A \neq \emptyset$). If $v \in A$ and $w \in {\cal V}_2$ satisfies $w \notin \Gamma_G(A)$, 
then $(v,w) \notin {\cal E}$. Hence
\begin{equation}
A \times \left( {\cal V}_2 \setminus \Gamma_G(A) \right) \subset
\left( {\cal V}_1 \times {\cal V}_2 \right) \setminus {\cal E}.
\end{equation}
Therefore
\begin{equation}
\left( \# A \right) \left( n - \# \Gamma_G(A) \right) \leq 
n^2 - \# {\cal E} \leq n - 1.
\end{equation}
If follows that
\begin{eqnarray}
\# \Gamma_G(A) & \geq & n - \frac{n-1}{\# A} = (\# A - 1) \frac{n}{\# A} + \frac{1}{\# A} \geq 
\# A - 1 + \frac{1}{\# A} \nonumber \\ 
& > & \#A - 1. 
\end{eqnarray}
Consequently Hall's condition $\# \Gamma_G(A) \geq \# A$ holds. \qed
\par
\medskip
Based on the above preparations, we prove {\bf Theorem 3.5}.
\par
\medskip
\noindent
{\bf Proof of Theorem 3.5}
\par
\medskip
\noindent
Let us show that the condition (3) of {\bf Theorem 3.4} holds. We first suppose that 
the bipartite graph $G = ({\cal V}^R,{\cal V}^C,{\cal E})$ has no perfect matchings. 
According to Hall's marriage theorem, $A \subset {\cal V}^R$ exits, 
so that $\# A > \# \Gamma_G(A)$. Then
\begin{equation}
{\cal K} = \{ j | v^R_j \in A \ {\rm and} \ v^C_j \notin \Gamma_G(A) \}
\end{equation}
is not empty and we can take $k \in {\cal K}$. The definition of ${\cal K}$ means
\begin{equation}
\label{k1}
\langle k,\ell \rangle \in {\cal E} \Longrightarrow v_\ell \in \Gamma_G(A)
\end{equation}
and
\begin{equation}
\label{k2}
\langle j,k \rangle \in {\cal E} \Longrightarrow v_j \in {\cal V}^R \setminus A.
\end{equation}
On the other hand, it follows from
 \begin{equation}
 {\cal E}[{\cal I}_N \setminus \{ k \}] = \{ \langle j, \ell \rangle | j \neq k \ {\rm and} \ \ell \neq k \}
 \end{equation}
that
\begin{equation}
\label{esetminus}
{\cal E} \setminus  {\cal E}[{\cal I}_N \setminus \{ k \}] = \{ \langle j,\ell \rangle | 
j = k \ {\rm or} \ \ell = k \}.
\end{equation}
Putting (\ref{k1}), (\ref{k2}) and (\ref{esetminus}) together, we obtain
\begin{equation}
{\cal E} \setminus  {\cal E}[{\cal I}_N \setminus \{ k \}] \subset 
\{ \langle j,k \rangle | v_j \in {\cal V}^R \setminus A \} \cup 
\{ \langle k,\ell \rangle | v_\ell \in \Gamma_G(A) \},
\end{equation}
which yields
\begin{equation}
\# {\cal E} - \# {\cal E}[{\cal I}_N \setminus \{ k \}] \leq 
N - \# A + \# \Gamma_G(A) \leq N-1.
\end{equation}
Let us now suppose that $\# {\cal E} = \# \Lambda \geq N^2 - 2 N + 2$. Then one can see that
\begin{equation}
 \# {\cal E}[{\cal I}_N \setminus \{ k \}] \geq \# {\cal E} - N + 1 \geq N^2 - 3 N + 3 
 = (N-1)^2 - (N-1) + 1.
\end{equation}
On account of {\bf Lemma 3.7}, $G[{\cal I}_N  \setminus \{ k \} ]$ has a perfect marching.
Therefore the condition (3) of {\bf Theorem 3.4} holds. \qed
\par
The following corollary of {\bf Theorem 3.5} gives the maximum number of continuous random 
matrix entries, when we have eigenvalue degeneracy.
\par
\medskip
\noindent
{\bf Corollary 3.8}
\par
\medskip
\noindent
We set $K = \mathbb{R}$ or $K=\mathbb{C}$. Suppose that $N \geq 2$ and let $X_{j \ell}: \Omega \rightarrow K$ ($j,\ell = 1,2, 
\cdots, N$) be independent random variables. We consider an $N \times N$ square 
random matrix $(X_{j \ell})_{j,\ell = 1,2,\cdots,N}$. If more than or equal to $N^2-2 N + 2$ of 
the random variables $X_{j \ell}$ are (broad) continuous, the random matrix $(X_{j \ell})$ has 
$N$ different eigenvalues with probability one. 
\par
\medskip
\noindent
{\bf Proof}
\par
\medskip
\noindent
It immediately follows from {\bf Corollary 2.9} and {\bf Theorem 3.5}.

\subsection{Symmetric matrices}

Next we consider symmetric matrices and corresponding graphs. 
In order to treat symmetric matrices, we introduce
\begin{equation}
{\cal I}^{(2)}_N = \{ \{j, \ell \} | j,\ell \in {\cal I}_N \}.
\end{equation}
Moreover, for ${\bf e} = \{j,\ell\} \in {\cal I}^{(2)}_N$, we define
\begin{equation}
E_{\bf e} = E_{\{j,\ell\}} = \frac{1}{2} ( E_{j \ell} + E_{\ell j}).
\end{equation}
\par
\medskip
\noindent
{\bf Lemma 3.9}
\par
\medskip
\noindent
Let us take an element $\sigma$ of the $N$-th order symmetric group $\mathfrak{S}_N$ 
and introduce the notations
\begin{equation}
{\bf S} = \left\{ \{ j, \sigma(j) \} | j \in {\cal I}_N \right\}
\end{equation}
and 
\begin{equation}
\varphi(x_{\bf S}) = \sum_{{\bf e} \in {\bf S}} x_{\bf e} E_{\bf e}.
\end{equation}
Suppose that $K$ is $\mathbb{R}$ or $\mathbb{C}$. Then,  for an $x_S \in K^{\# {\bf S}}$, the eigenvalues of 
$\varphi$ do not degenerate, and are all different from zero.
\par
\medskip
\noindent
{\bf Proof}
\par
\medskip
\noindent
In the same way as the proof of {\bf Lemma 3.1}, by decomposing $\sigma$ into a product of cyclic 
permutations, we find that it is sufficient to show this lemma when $\sigma$ is a cyclic permutation 
$(1 2 \cdots N)$. Let us restrict ourselves to that case.
\par
If  $N \leq 2$, ${\bf S}$ consists of a unique element $\{ 1, \sigma(1) \}$. Then the required result 
holds when $x_{\{1,\sigma(1)\}} \neq 0$. In the following, we assume $N \geq 3$. Then 
the number of the elements of ${\bf S}$ is $N$. We write the variable $x_{\{ j,\sigma(j)\}}$ as $x_j$. 
Let us put $x_1 = x_2 = \cdots = x_{N-1}=1$ and $x_N = \epsilon$, and consider the matrix 
$P(\epsilon) = \varphi(1,\cdots,1,\epsilon)$. When $\epsilon = 0$, $P(0)$ is a tridiagonal matrix 
and the eigenvalues are $\cos\left( k\pi/(N+1 ) \right)$ ($k=1,2,\cdots,N)$. 
The required result evidently 
holds for an even $N$. In the following we consider the odd $N$ case. As the eigenvalues of 
$P(0)$ do not degenerate, the eigenvalues of $P(\epsilon)$ do not degenerate either, provided 
that $|\epsilon|$ is small enough, owing to the continuity of eigenvalues. On the other hand, 
since $\det P(\epsilon) = \epsilon/2^N$, all the eigenvalues are non-zero, if $\epsilon \neq 0$. 
Therefore the require result holds, when $|\epsilon|$ has a non-zero but sufficiently small value. \qed
\par
\medskip
By using {\bf Lemma 3.9}, similarly as {\bf Lemma 3.3} and {\bf Theorem 3.4}, we can prove the  following 
lemma and theorem for symmetric matrices.
\par
\medskip
\noindent
{\bf Lemma 3.10}
\par
\medskip
\noindent
Let us suppose that $\mathfrak{L} \subset {\cal I}^{(2)}_N$ and $\varphi(x_{\mathfrak{L}}) = 
\sum_{{\bf e} \in \mathfrak{L}} x_{\bf e} E_{\bf e}$ with $x_{\bf e} \in K$ ($K$ is $\mathbb{R}$ or $\mathbb{C}$). Moreover we define
\begin{equation}
\mathfrak{E} = \Big\{ \langle j,\ell \rangle \in {\cal V}^R \times {\cal V}^C \Big| \{ j,\ell \} \in \mathfrak{L} \Big\}.
\end{equation}
If a bipartite graph 
$G = ({\cal V}^R,{\cal V}^C, \mathfrak{E})$ has a perfect matching, then an $x_{\mathfrak{L}}$ exits, so that all the 
$N$ eigenvalues of $\varphi(x_{\mathfrak{L}})$ are different from each other and all non-zero. 
\par
\medskip
\noindent
Let $S$ be a subset of $ {\cal I}_N$. Corresponding to a bipartite graph $G = ({\cal V}^R,{\cal V}^C, \mathfrak{E})$, 
a bipartite subgraph $G[S] = \left( {\cal V}^R[S],{\cal V}^C[S],\mathfrak{E}[S] \right)$ is defined as
\begin{equation}
{\cal V}^R[S] = \{ v_j^R | j \in S \}, \ \ \ 
{\cal V}^C[S] = \{ v_j^C | j \in S \}, \ \ \ 
\mathfrak{E}[S] = \{ \langle j,\ell \rangle \in \mathfrak{E} | j,\ell \in S \}.
\end{equation}
\par
\medskip
\noindent
{\bf Theorem 3.11}
\par
\medskip
\noindent
Let us suppose that $\mathfrak{L} \subset {\cal I}^{(2)}_N$ and  
$\varphi(x_{\mathfrak{L}}) = \sum_{{\bf e} \in {\cal W}} x_{\bf e} E_{\bf e}$ with $x_{\bf e} \in K$ ($K$ is $\mathbb{R}$ or $\mathbb{C}$).
Then the following conditions are equivalent.
\par
\medskip
\noindent
(1) An $x_{\mathfrak{L}}$ exists, so that the $N$ eigenvalues of $\varphi(x_{\mathfrak{L}})$ are different from each other. 
\par
\medskip
\noindent
(2) An $x_{\mathfrak{L}}$ exists, so that $\varphi(x_{\mathfrak{L}})$ has no zero eigenvalues or a simple zero eigenvalue.
\par
\medskip
\noindent
(3) The bipartite graph $G = ({\cal V}^R,{\cal V}^C,\mathfrak{E})$ has a perfect matching, or 
$j \in {\cal I}_N$ exists and the bipartite subgraph $G[{\cal I}_N \setminus \{j\}]$ has 
a perfect matching. 

\section{Eigenvalue degeneracy probability for random matrices}
\setcounter{equation}{0}
\renewcommand{\theequation}{4.\arabic{equation}}

Let $K = \mathbb{R}$ or $K = \mathbb{C}$. We suppose that $Z_{j \ell}$ ($j,\ell=1,2,\cdots,N)$ 
are independent and (broad) continuous $K$-valued random variables. Moreover we introduce 
independent random variables $Y_{j \ell}$ ($j,\ell = 1,2,\cdots,N$) obeying Bernoulli distribution 
with a parameter $p$. A random variable obeying Bernoulli distribution with a parameter $p$ 
takes a value $1$ with probability $p$ and another value $0$ with probability $1 - p$. 
\par
We consider the probability for the random matrix $(Y_{j \ell} Z_{j \ell} )_{j,\ell=1,2,\cdots,N}$ to 
have $N$ distinct eigenvalues. First let us arbitrarily fix $y_{j \ell} \in \{ 0,1 \}$ ($j,\ell = 1,2,\cdots,N$) 
and evaluate the conditional probability under the condition $Y_{j \ell} = y_{j \ell}$. 
It follows from ${\bf Theorem 2.5}$ that this conditional probability is zero or one, and 
the necessary and sufficient condition to have the conditional probability one is:
\par
\medskip
\noindent
there exits at least one matrix $(z_{j \ell})_{j,\ell=1,2,\cdots,N}$ with $z_{j \ell} \in K$, so that
 the matrix $(y_{j \ell} z_{j \ell})_{j,\ell=1,2,\cdots,N}$ has $N$ distinct eigenvalues.
\par
\medskip
\noindent
This conditional probability does not depend on the distributions of $Z_{j \ell}$, 
provided that they are independent and (broad) continuous. Therefore the probability 
for the matrix $(Y_{j \ell} Z_{j \ell})_{j,\ell=1,2,\cdots,N}$ to have $N$ distinct eigenvalues 
does not depend on the distributions of $Z_{j \ell}$ either. It turns out to depend only 
on $N$ and $p$. Moreover, we can readily see from {\bf Theorem 3.4} that this 
probability is equal to the probability for the random bipartite graph $G=({\cal V}^R, {\cal V}^C, {\cal E})$, 
${\cal E} = \{\langle j,\ell \rangle | Y_{j \ell}= 1 \}$ to satisfy the following condition.
\par
\medskip
\noindent
{\bf Condition 4.1}
\par
\medskip
\noindent
The bipartite graph $G = ({\cal V}^R,{\cal V}^C,{\cal E})$ has a perfect matching, or 
$j \in {\cal I}_N$ exists and the bipartite subgraph $G[{\cal I}_N \setminus \{j\}]$ has 
a perfect matching. Here ${\cal I}_N = \{1,2,\cdots,N \}$.
\par
\par
\medskip
\noindent
Thus, as a simpler problem, let us first consider the probability for a random graph to 
have a perfect matching. For that problem, there is a known result explained below 
as {\bf Proposition 4.2}.
\par
\medskip
\noindent
{\bf Remark}
\par
\medskip
\noindent
Another consequence of {\bf Theorem 3.4} is that the probability for the matrix 
$(Y_{j \ell} Z_{j \ell})_{j,\ell=1,2,\cdots,N}$ to have $N$ distinct eigenvalues is 
equal to the probability that it has no zero eigenvalues or a simple zero eigenvalue. 
On the other hand, if an $N \times N$  matrix has $N$ distinct eigenvalues, then 
in general it has no zero eigenvalues or a simple zero eigenvalue. Therefore, if 
the eigenvalues of $(Y_{j \ell} Z_{j \ell})_{j,\ell=1,2,\cdots,N}$ degenerate, then zero 
eigenvalues exist and degenerate with probability one. Namely, as for this random matrix 
model, it can be said that accumulation of the eigenvalues to the origin causes 
eigenvalue degeneracy.

\subsection{The probability for a random graph to have a perfect matching}

In this subsection, we introduce the following proposition for the asymptotic probability for 
a random bipartite graph to have a perfect matching. This proposition was originally proved 
by Erd\"os and R\'enyi\cite{ER}. It is useful in the study of asymptotic degeneracy 
probability for the eigenvalues of the corresponding random matrices, 
\par
\medskip
\noindent
{\bf Proposition 4.2}
\par
\medskip
\noindent
We consider a random bipartite graph $G_n = ( {\cal V}_1,{\cal V}_2,{\cal E})$ with 
an edge set ${\cal E} \subset {\cal V}_1 \times {\cal V}_2$ and $\# {\cal V}_1 = \# {\cal V}_2 = n$.
Here vertex sets are
\begin{equation}
{\cal V}_1 = \{ v_1v_2,\cdots,v_n \}, \ \ \ {\cal V}_2 = \{ w_1,w_2,\cdots,w_n \}
\end{equation}
and $(v_j,w_\ell) \in {\cal V}_1 \times {\cal V}_2$ is abbreviated as $\langle j,\ell \rangle$. 
Each $\langle j,\ell \rangle$ 
($j,\ell = 1,2,\cdots,n$) independently satisfies $\langle j,\ell \rangle \in {\cal E}$ with probability $p$. 
Suppose that the probability parameter $p$ is dependent on $n$ and written as
\begin{equation}
p = p(n) = \frac{\log n + c}{n} + o\left( \frac{1}{n} \right), \ \ \ n \rightarrow \infty
\end{equation}
with a constant $c \in \mathbb{R}$. Then the probability $\mathbb{P}$ for the bipartite graph $G_n$  to have
a perfect matching is asymptotically evaluated as
\begin{equation}
\lim_{n \rightarrow \infty} \mathbb{P}(G_n \ {\rm has} \  {\rm a} \ {\rm  perfect} \ {\rm  matching}) = e^{-\lambda},
\end{equation}
where $\lambda = 2 e^{-c}$.
\par
\medskip
\noindent
In the remaining part of this subsection, we give a step-by-step proof of {\bf Proposition 4.2} in reference to 
Frieze and Karo\'nski's book\cite{FK}. Intermediate results are also useful in the study of eigenvalue 
degeneracy. Let us begin with a definition and a proposition.
\par
\medskip
\noindent
{\bf Definition 4.3}
\par
\medskip
\noindent
Suppose that $\mathbb{Z}_+ = \{ m \in \mathbb{Z} | m \geq 0 \}$ and $k \in \mathbb{Z}_+$. We 
define a $k$-th order polynomial $\displaystyle \binom{x}{k}$ 
with indeterminates $x$ as
\begin{equation}
\binom{x}{0} = 1 \ \  {\rm and} \ \ 
\binom{x}{k}  = 
\frac{x (x - 1) \cdots (x - k + 1)}{k!} \ {\rm for} \ k \geq 1.
\end{equation}
\par
\medskip
\noindent
When $x = n \in \mathbb{Z}+$, $\dbinom{n}{k} \geq 0$. 
In particular, if $n < k$, we have $\dbinom{n}{k} = 0$.
\par
\medskip
\noindent
{\bf Proposition 4.4}
\par
\medskip
\noindent
For $k,m,n \in \mathbb{Z}_+$, the following equality holds.
\begin{equation}
\sum_{k=0}^m (-1)^k \binom{n+1}{k} = (-1)^m \binom{n}{m}.
\end{equation}
\par
\medskip
\noindent
{\bf Proof}
\par
\medskip
\noindent
Mathematical induction with respect to $m$ is employed. For $m=0$, both sides of the 
equality is $1$. Let us assume that the equality holds for general $m \in \mathbb{Z}_+$.  
When $m$ is replaced with $m+1$, we have
\begin{eqnarray}
& & \sum_{k=0}^{m+1} (-1)^k \binom{n+1}{k} = \sum_{k=0}^m (-1)^k \binom{n+1}{k} + (-1)^{m+1} 
\binom{n+1}{m+1} \nonumber \\ 
& = & (-1)^m \binom{n}{m} + (-1)^{m+1} \binom{n+1}{m+1} =  (-1)^{m+1} \binom{n}{m+1}.
\end{eqnarray}
The assumption is used in the derivation of the second line. \qed
\par
\medskip
\noindent
In the following we denote an expectation value of a random variable $X$ by $\mathbb{E} X$.
\par
\medskip
\noindent
{\bf Proposition 4.5}
\par
\medskip
\noindent
Suppose that a $\mathbb{Z}_+$ valued random variable $X$ satisfies 
\begin{equation}
\mathbb{E} \binom{X}{k} = \beta_k < \infty
\end{equation}
for arbitrary $k \in \mathbb{Z}_+$. Then the following inequality holds for 
arbitrary $j, \ell \in \mathbb{Z}_+$.
\begin{equation}
\sum_{k=j}^{j + 2 \ell + 1} (-1)^{k - j} \binom{k}{j} \beta_k \leq 
\mathbb{P}(X = j) \leq \sum_{k=j}^{j + 2 \ell} (-1)^{k - j} \binom{k}{j} \beta_k.
\end{equation}
\par
\medskip
\noindent
{\bf Proof}
\par
\medskip
\noindent
Since
\begin{equation}
\beta_k = \sum_{n=0}^\infty \binom{n}{k} \mathbb{P}(X=n),
\end{equation}
we obtain
\begin{eqnarray}
\sum_{k=j}^{j + m} (-1)^{k - j} \binom{k}{j} \beta_k  
& = & \sum_{k=j}^{j+m} (-1)^{k-j} \binom{k}{j} \sum_{n=0}^\infty \binom{n}{k} \mathbb{P}(X=n) 
\nonumber \\ & = & 
\sum_{n=0}^\infty \mathbb{P}(X=n) \sum_{k=j}^{j+m} (-1)^{k-j} \binom{k}{j} \binom{n}{k}
\end{eqnarray}
for $j,m \in \mathbb{Z}_+$. Using
\begin{equation}
\binom{k}{j} \binom{n}{k} = \binom{n}{j} \binom{n - j}{k-j}
\end{equation}
and {\bf Proposition 4.4}, we further find
\begin{eqnarray}
& & \sum_{k=j}^{j + m} (-1)^{k - j} \binom{k}{j} \beta_k  
= \sum_{n=j}^\infty \mathbb{P}(X=n)  \binom{n}{j} \sum_{k=j}^{j+m} (-1)^{k-j}\binom{n-j}{k-j}
\nonumber \\ 
& = & \mathbb{P}(X=j) + (-1)^m \sum_{n=j+1}^\infty \mathbb{P}(X=n)  \binom{n}{j} \binom{n-j-1}{m}.
\end{eqnarray}
Considering the cases $m$ odd and $m$ even, respectively, we arrive at the required inequality. \qed
\par
\medskip
\noindent
{\bf Proposition 4.6}
\par
\medskip
\noindent
Let $\{ X_n \}_{n \in \mathbb{N}}$ be a sequence of $\mathbb{Z}_+$ valued random variables, 
and for arbitrary $k \in \mathbb{Z}_+$ assume that  $\displaystyle \lim_{n \rightarrow \infty} \mathbb{E}\binom{X_n}{k}$ 
converges to a finite value $\beta_k$. Then, if a series
\begin{equation}
\sum_{k=j}^\infty (-1)^{k - j} \binom{k}{j} \beta_k
\end{equation}
converges, $\displaystyle \lim_{n \rightarrow \infty} \mathbb{P}(X_n = j)$ also converges and 
\begin{equation}
\lim_{n \rightarrow \infty} \mathbb{P}(X_n = j) = \sum_{k=j}^\infty  (-1)^{k-j} \binom{k}{j} \beta_k
\end{equation}
holds.
\par
\medskip
\noindent
{\bf Proof}
\par
\medskip
\noindent
Let us put $\displaystyle \beta_k^{(n)} = \mathbb{E}\binom{X_n}{k}$. Since $\displaystyle 
\lim_{n \rightarrow \infty} \beta_k^{(n)} < \infty$, it can be seen that $\beta_k^{(n)} < \infty$ for 
sufficiently large $n$. Therefore, for such $n$, 
\begin{equation}
\sum_{k=j}^{j + 2 \ell + 1} (-1)^{k - j} \binom{k}{j} \beta_k^{(n)} \leq 
\mathbb{P}(X_n = j) \leq \sum_{k=j}^{j + 2 \ell} (-1)^{k - j} \binom{k}{j} \beta_k^{(n)}
\end{equation}
holds due to {\bf Proposition 4.5}. Taking the limit $n \rightarrow \infty$, we find
\begin{eqnarray}
& & \sum_{k=j}^{j + 2 \ell + 1} (-1)^{k - j} \binom{k}{j} \beta_k \leq \liminf_{n \rightarrow \infty}
\mathbb{P}(X_n = j) \nonumber \\ 
& \leq & \limsup_{n \rightarrow \infty} \mathbb{P}(X_n = j) \leq \sum_{k=j}^{j + 2 \ell} (-1)^{k - j} 
\binom{k}{j} \beta_k.
\end{eqnarray}
We finally take the limit $\ell \rightarrow \infty$ to obtain the required result. \qed
\par
\medskip
\noindent
Let us now introduce a formula giving an evaluation of  $\displaystyle \mathbb{E}\binom{X}{k}$ for a 
special $\mathbb{Z}_+$ valued random variable $X$.
\par
\medskip
\noindent
{\bf Proposition 4.7}
\par
\medskip
\noindent
Suppose that $B_1,B_2,\cdots,B_m$ are $\{0,1\}$-valued random variables. When a $\mathbb{Z}_+$-valued 
random variable $X$ is defined as
\begin{equation}
X = \sum_{j=1}^m  B_j, 
\end{equation}
an equality
\begin{equation}
\mathbb{E}\binom{X}{k} = \sum_{\substack{I \subset S \\ \# I = k}} \mathbb{E} \left\{ 
\prod_{j \in I} B_j \right \}
\end{equation}
holds for $S=\{1,2,\cdots,m\}$ and $k \in \mathbb{N}$.
\par
\medskip
\noindent
{\bf Proof}
\par
\medskip
\noindent
Let us first prove an identity
\begin{equation}
\label{identity}
\prod_{j \in I} B_j = \sum_{\substack{J \subset S \\ J \supset I}} \prod_{j \in J} B_j \prod_{\ell \in S \setminus J} ( 1 - B_\ell)
\end{equation}
for $I \subset S$. If $\displaystyle \prod_{j \in I} B_j = 0$, the both sides of (\ref{identity}) are 
obviously zero. If $\displaystyle \prod_{j \in I} B_j = 1$, each term in the sum over $J$ in the 
right hand side is zero, unless a condition $B_j = 1$  ($j \in J$) and $B_\ell = 0$ ($\ell \in S \setminus J$) is 
satisfied. Only a single term satisfies such a condition and its value is one. Therefore the both sides 
of (\ref{identity}) are one. 
\par
Using the identity (\ref{identity}), one can see that
\begin{eqnarray}
\sum_{\substack{I \subset S \\ \# I = k}} \mathbb{E} \left\{ \prod_{j \in I} B_j \right \} & = & 
\sum_{\substack{I \subset S \\ \# I = k}} \sum_{\substack{J \subset S \\ J \supset I}} \mathbb{E} \left\{ 
\prod_{j \in J} B_j \prod_{\ell \in S \setminus J} ( 1 - B_\ell) \right\}
\nonumber \\ 
& = & \sum_{\substack{J \subset S \\ \# J \geq k}} \sum_{\substack{I \subset J \\ \# I = k}} 
\mathbb{E} \left\{ \prod_{j \in J} B_j \prod_{\ell \in S \setminus J} ( 1 - B_\ell) \right\} \nonumber \\ 
& = & \sum_{n=k}^m \sum_{\substack{J \subset S \\ \# J = n}} \binom{n}{k} 
\mathbb{E} \left\{ \prod_{j \in J} B_j \prod_{\ell \in S \setminus J} ( 1 - B_\ell) \right\}.
\end{eqnarray}
On the other hand, we can write the probability $\mathbb{P}(X = x)$ as
\begin{equation}
\mathbb{P}(X = n) = \sum_{\substack{J \subset S \\ \# J = n}} \mathbb{E} \left\{ \prod_{j \in J} B_j \prod_{\ell \in S \setminus J} ( 1 - B_\ell) \right\}.
\end{equation}
Therefore we find
\begin{equation}
\sum_{\substack{I \subset S \\ \# I = k}} \mathbb{E} \left\{ \prod_{j \in I} B_j \right \} 
= \sum_{n=k}^m  \binom{n}{k} \mathbb{P}(X = n) = 
\mathbb{E}\binom{X}{k},
\end{equation}
which gives the required result. \qed
\par
\medskip
\noindent
A necessary condition for a bipartite graph $G =( {\cal V}_1,{\cal V}_2,{\cal E})$ with $\# {\cal V}_1 = 
\# {\cal V}_2 = n$ to have a perfect matching is the absence of isolated points defined below. 
\par
\medskip
\noindent
{\bf Definition 4.8}
\par
\medskip
\noindent
In a bipartite graph $G = ({\cal V}_1,{\cal V}_2,{\cal E})$, \ $v \in {\cal V}_1$ satisfying 
\par
\medskip
\noindent
$(v,w) \notin {\cal E} \ {\rm for} \ {\rm arbitrary} \ w \in {\cal V}_2$
\par
\medskip
\noindent
and $w \in {\cal V}_2$ satisfying
\par
\medskip
\noindent
$(v,w) \notin {\cal E} \ {\rm for} \ {\rm arbitrary} \ v \in {\cal V}_1$
\par
\medskip
\noindent
are called isolated points of $G$.
\par
\medskip
\noindent
We first evaluate the probability to have no isolated points, because it will eventually 
turn out to dominate the probability to have a perfect matching in the limit $n \rightarrow \infty$. 
\par
\medskip
\noindent
{\bf Proposition 4.9}
\par
\medskip
\noindent
Let us come back to the notations of {\bf Proposition 4.2} and again consider a random bipartite graph $G_n = ({\cal V}_1,{\cal V}_2,{\cal E})$ with ${\cal E} 
\subset {\cal V}_1 \times {\cal V}_2$ and $\# {\cal V}_1 = \# {\cal V}_2 = n$. 
Each $\langle j,\ell \rangle \in {\cal V}_1 \times {\cal V}_2$ 
($j,\ell=1,2,\cdots,n$) independently satisfies $\langle j,\ell \rangle \in {\cal E}$ with probability $p$. The 
probability parameter $p$ depends on $n$ as
\begin{equation}
\label{parameter}
p = p(n) = \frac{\log n + c}{n} + o\left( \frac{1}{n} \right), \ \ \ n \rightarrow \infty, 
\end{equation} 
where $c \in \mathbb{R}$ is a constant. Suppose that the number of the isolated points 
of the random bipartite graph 
$G_n = ( {\cal V}_1,{\cal V}_2,{\cal E})$ is $X_n$. The probability to have $X_n= x$ for a fixed 
$x \in \mathbb{Z}_+$ is asymptotically evaluated as
\begin{equation}
\lim_{n \rightarrow \infty} \mathbb{P}(X_n = x) = e^{-\lambda} \frac{\lambda^x}{x!}
\end{equation}
with $\lambda = 2 e^{-c}$.
\par
\medskip
\noindent
{\bf Proof}
\par
\medskip
\noindent
It follows from {\bf Proposition 4.7} that
\begin{equation}
\mathbb{E}\binom{X_n}{k} = \sum_{\substack{I \subset {\cal V}_1 \cup {\cal V}_2 \\ \# I = k}} 
\mathbb{P}\left( {\rm the} \ {\rm elements } \ {\rm of} \  I \ 
{\rm are} \ {\rm  all} \ {\rm isolated} \ {\rm  points} \ {\rm of} \  G_n \right).
\end{equation}
Let us write $I \subset {\cal V}_1 \cup {\cal V}_2$ with $\# I = k$ as
\begin{equation}
I = I_1 \cup I_2, \ \ \ I_1 \subset {\cal V}_1, \ \ \ I_2 \subset {\cal V}_2
\end{equation}with
\begin{equation}
\# I_1 = k_1 \geq 0, \ \ \ \# I_2 =k_2 \geq 0, \ \ \ k_1 + k_2 = k.
\end{equation}
There are $\displaystyle \binom{n}{k_1} \binom{n}{k_2}$ ways to choose such $\{ I_1, I_2 \}$. 
Then the probabilities to have $(v,w) \notin {\cal E}$ for arbitrary $(v,w) \in I_1 \times I_2$, 
for arbitrary $(v,w) \in I_1 \times \left( {\cal V}_2 \setminus I_2 \right) $ and for arbitrary $(v,w) 
\in \left( {\cal V}_1 \setminus I_1 \right) \times  I_2$ are
\begin{equation}
(1 - p)^{k_1 k_2},  (1 - p)^{k_1 (n - k_2)} \ {\rm and} \ (1 - p)^{k_2 (n - k_1)},
\end{equation}
respectively. Hence we obtain
\begin{eqnarray}
\mathbb{E}\binom{X_n}{k} & =& \sum_{k_1+k_2 = k} \binom{n}{k_1} \binom{n}{k_2} 
(1 - p)^{k_1 k_2} (1 - p)^{k_1 (n - k_2)} (1 - p)^{k_2 (n - k_1)} \nonumber \\ 
& =& \sum_{k_1+k_2 = k} \binom{n}{k_1} \binom{n}{k_2} 
(1 - p)^{n k - k_1 k_2}.
\end{eqnarray}
One can see from (\ref{parameter}) that
\begin{equation}
\log(1 - p) = - \frac{\log n + c}{n} + o\left( \frac{1}{n} \right),
\end{equation}
which leads to 
\begin{eqnarray}
\mathbb{E}\binom{X_n}{k} & =& \sum_{k_1+k_2 = k} \frac{n^k ( 1 + o(1))}{k_1! k_2!} 
{\rm exp}\left\{ (n k - k_1 k_2)  \log(1 - p) \right\} \nonumber \\  
& = &  \sum_{k_1+k_2 = k} \frac{1 + o(1)}{k_1! k_2!} 
{\rm exp}\left\{ k \log n + n k \log(1 - p) + o(1) \right\} \nonumber \\ 
& = &  \frac{1}{k!} \sum_{k_1=0}^k \binom{k}{k_1} {\rm exp}\left( - c k \right) + o(1) \nonumber \\ 
& = &  \frac{2^k e^{-c k}}{k!}  + o(1) = \frac{\lambda^k}{k!} + o(1).
\end{eqnarray}
Now we can utilize {\bf Proposition 4.6} to find
\begin{eqnarray}
\lim_{n \rightarrow \infty} \mathbb{P}(X_n = x) & = & \sum_{k=x}^{\infty} (-1)^{k-x} \binom{k}{x} \frac{\lambda^k}{k!} = \sum_{k=x}^{\infty}  \frac{(-1)^{k-x} \lambda^k}{(k-x)! x!} \nonumber \\ 
& = & \frac{\lambda^x}{x!} \sum_{\ell = 0} \frac{(-\lambda)^\ell}{\ell!} = \frac{\lambda^x}{x!} e^{-\lambda}.
\end{eqnarray}
\qed
\par
\medskip
\noindent
In order to prove {\bf Proposition 4.2}, we need to show that the probability for a random bipartite graph 
$G$ to have neither isolated points nor perfect matchings converges to zero. 
\par
When a bipartite graph $G = ({\cal V}_1,{\cal V}_2,{\cal E})$ has no perfect matchings, 
{\bf Proposition 3.6} (Hall's marriage theorem) means that
\par
\medskip
\noindent
there exists $A_1 \subset {\cal V}_1$ satisfying $\# \Gamma_G(A_1) \leq \# A_1 - 1 $. 
\par
\medskip
\noindent
The same proposition can also be applied to the case when ${\cal V}_1$ and ${\cal V}_2$ are interchanged 
and consequently
\par
\medskip
\noindent
there exists $A_2 \subset {\cal V}_2$ satisfying $\# \Gamma_G(A_2) \leq \# A_2 - 1 $. 
\par
\medskip
\noindent
As for such sets $A_1$ and $A_2$, we have the following proposition.
\par
\medskip
\noindent
{\bf Proposition 4.10}
\par
\medskip
\noindent
Let us consider a bipartite graph $G = ({\cal V}_1,{\cal V}_2,{\cal E})$ with ${\cal V}_1 = {\cal V}_2 = n$ 
and suppose that $G$ has no perfect matchings. Among the sets $A \subset {\cal V}_1$ and $A \subset 
{\cal V}_2$ which satisfy $\# \Gamma_G(A) \leq \# A - 1$, we choose one set minimizing $\# A$ and 
call it $I$. Then $I$ satisfies the followings.
\par
\medskip
\noindent
(1) $\# \Gamma_G(I) = \# I - 1$.
\par
\medskip
\noindent
(2) $ \# I \leq (n + 1)/2$.
\par
\medskip
\noindent
(3) $\# (\Gamma_G(w) \cap I) \geq 2$ for arbitrary $w \in \Gamma_G(I)$. 
\par
\medskip
\noindent
{\bf Proof}
\par
\medskip
\noindent
(1) We arbitrarily choose $v \in I$ and put $I' = I \setminus \{ v \}$. As $I$ minimizes the number 
of elements in the family of sets unsatisfying Hall's condition, one finds $\# \Gamma_G(I') \geq 
\# I'$. 
Moreover, as $I' \subset I$, we obtain $\Gamma_G(I') \subset \Gamma_G(I)$. It follows that
\begin{equation}
\# \Gamma_G(I) \geq \# \Gamma_G(I') \geq \# I' = \# I - 1.
\end{equation}
On the other hand, since $I$ does not satisfy Hall's condition, one sees $\# \Gamma_G(I) \leq \# I - 1$. 
Therefore the required result together with the equality 
$\Gamma_G(I) = \Gamma_G(I')$ holds.
\par
\medskip
\noindent
(2) Let us first assume $I \subset {\cal V}_1$. If $\# I > (n+1)/2$, then $\# \Gamma_G(I) = \# I  - 1 
> (n-1)/2$ from (1). We now put $K = {\cal V}_2 \setminus \Gamma_G(I)$ and find 
$\# K= n -  \# \Gamma_G(I) < (n+1)/2$. On the other hand, as $\Gamma_G(K) \subset {\cal V}_1 
\setminus I$, 
\begin{equation}
\# \Gamma_G(K) \leq n - \# I = n - \# \Gamma_G(I) - 1 = \# K - 1.
\end{equation}
This is contradictory because $K$ does not satisfy Hall's condition in spite of the fact that $\# K < (n+1)/2 < \# I$.
A similar argument holds in the other case $I \subset {\cal V}_2$.
\par
\medskip
\noindent
(3) We arbitrarily choose $w \in \Gamma_G(I)$. From the definition of $\Gamma_G$, there exists 
$v \in I$ and $(v,w) \in {\cal E}$. Now we define $I'= I \setminus \{ v \}$, then $\Gamma_G(I) = \Gamma_G(I')$ 
from (1). Hence $v' \in I'$ exists, so that $(v',w) \in {\cal E}$. Therefore $\# (\Gamma_G(w) \cap I)  \geq 2$. \qed
\par
\medskip
\noindent
Due to {\bf Proposition 4.10}, if a bipartite graph $G$ has no perfect matchings, then it has at least one isolated point or satisfies the following condition.
\par
\medskip
\noindent
{\bf Condition 4.11}
\par
\medskip
\noindent
There exists $k \in \mathbb{N}$ ($2 \leq k \leq (n+1)/2$) satisfying 
$I \subset {\cal V}_1$, $J \subset {\cal V}_2$ (or $I \subset {\cal V}_2$,  
$J \subset {\cal V}_1$) with $\# I = k$, $\# J = k-1$, so that $\Gamma_G(I) = J$ and 
$\forall w \in J$, $\# (\Gamma_G(w) \cap I )\geq 2$.
\par
\medskip
\noindent
The case $k=1$ corresponds to graphs with at least one isolated point. Now 
we asymptotically evaluate the probability that {\bf Condition 4.11} holds.
\par
\medskip
\noindent
{\bf Proposition 4.12}
\par
\medskip
\noindent
Let us suppose  $\# {\cal V}_1 = \# {\cal V}_2 = n$ and consider a
random bipartite graph $G_n({\cal V}_1,{\cal V}_2,{\cal E})$. 
Each $\langle j,\ell \rangle \in {\cal V}_1 \times {\cal V}_2$ 
($j,\ell=1,2,\cdots,n$) independently satisfies $\langle j,\ell \rangle \in {\cal E}$ with probability $p$. The 
probability parameter $p$ depends on $n$ as
\begin{equation}
p = p(n) = \frac{\log n + c}{n} + o\left( \frac{1}{n} \right), \ \ \ n \rightarrow \infty
\end{equation} 
with a constant $c \in \mathbb{R}$. Then
\begin{equation}
\lim_{n  \rightarrow \infty} \mathbb{P}\left(G_n \ {\rm  satisfies} \  {\bf Condition \ 4.11} \right) = 0
\end{equation}
holds.
\par
\medskip
\noindent
{\bf Proof}
\par
\medskip
\noindent
Let us first fix $k \in \mathbb{N}$ satisfying $2 \leq k \leq (n + 1)/2$. 
We choose vertex sets $I \subset {\cal V}_1$ and $J \subset {\cal V}_2$ with $\# I = k$ and $\# J = k - 1$. There are $\displaystyle \binom{n}{k} \binom{n}{k-1}$ ways to choose such $\{ I, J\}$. Then the probability to have $(v,w) \notin {\cal E}$ for all $(v,w) \in I \times {\cal V}_2 \setminus J$ is $\displaystyle (1-p)^{k(n - k + 1)}$. 
It follows that
\begin{equation}
\mathbb{P}(\Gamma_G(I) = J) \leq (1 - p)^{k(n - k + 1)}
\end{equation}
for each $\{ I,J \}$. 
\par
Next we fix a vertex $w \in J$ and choose two distinct vertices $v,v' \in I$. For each $w$ there are $\displaystyle \binom{k}{2}$ ways to choose such $v,v'$. The probability to have $(v,w) \in {\cal E}$ and 
$(v',w) \in {\cal E}$ for each triple $\{ w,v,v' \}$ is $\displaystyle p^2$. Thus we obtain
\begin{equation}
\mathbb{P}(\exists \{v,v'\}, (v,w) \in {\cal E} \ {\rm and} \ 
(v',w) \in {\cal E}) \leq \binom{k}{2} p^2
\end{equation}
for each $w$.
\par
We can see from the above argument that
\begin{eqnarray}
& & \mathbb{P}\left(G_n \ {\rm  satisfies} \  {\bf Condition \ 4.11} \right) \nonumber \\
& \leq & 2 
\sum_{k=2}^{\lfloor (n+1)/2 \rfloor} \binom{n}{k} \binom{n}{k-1} \left\{ \binom{k}{2} p^2 \right\}^{k-1} 
(1 - p)^{k(n - k + 1)} \nonumber \\ 
& \leq & 2 \sum_{k=2}^{\lfloor (n+1)/2 \rfloor} \left( \frac{n e}{k} \right)^k
\left( \frac{n e}{k - 1} \right)^{k-1}  \left\{ \binom{k}{2} p^2 \right\}^{k-1} 
e^{-p k(n - k + 1)}.
\end{eqnarray}
Note that there is a factor $2$ in front of the sum over $k$ due to another case $I \in {\cal V}_2$ and $J \in {\cal V}_1$. 
The first term with $k=2$ is
\begin{eqnarray}
& & \frac{(n e)^3 }{2} p^2 e^{-2 p (n - 1)} \nonumber \\ 
& = & \frac{(n e)^3 }{2}  \left\{ \frac{\log n + c}{n} + o\left( \frac{1}{n} \right) \right\}^2 
{\rm exp}\left[ -2  \left\{ \frac{\log n + c}{n} + o\left( \frac{1}{n} \right) \right\} (n - 1) \right] 
\nonumber \\ 
& = & O\left( \frac{(\log n)^2}{n} \right) \rightarrow 0, \ \ \ n \rightarrow \infty.
\end{eqnarray}
The sum of the other terms with $k \geq 3$ is
\begin{eqnarray}
\varOmega & =  & 2 \sum_{k=3}^{\lfloor (n+1)/2 \rfloor} \left( \frac{n e}{k} \right)^k
\left( \frac{n e}{k - 1} \right)^{k-1}  \left\{ \binom{k}{2} p^2 \right\}^{k-1} 
e^{-p k(n - k + 1)} \nonumber \\ 
& = & \frac{4}{n e p^2}  \sum_{k=3}^{\lfloor (n+1)/2 \rfloor} \frac{1}{k} 
\left( \frac{(n e p)^2}{2} \right)^k
e^{-p k(n - k + 1)}.
\end{eqnarray}
Here one can see from $k \leq (n + 1)/2$ that
\begin{equation}
e^{-p (n - k + 1)} \leq e^{-p ( n + 1)/2} \leq e^{-p n /2}.
\end{equation}
Thus we find
\begin{equation}
\label{k3a}
\varOmega \leq \frac{1}{n p^2}  \sum_{k=3}^{\lfloor (n+1)/2 \rfloor}  u^k \leq \frac{1}{n p^2} \frac{u^3}{1 - u},
\end{equation}
where
\begin{equation}
u = \frac{(n e p)^2}{2} e^{-p n /2}.
\end{equation}
Let us take constants $c_1,c_2 \in \mathbb{R}$ satisfying
\begin{equation}
\frac{\log n + c_1}{n} \leq p(n) \leq \frac{\log n + c_2}{n}
\end{equation}
for sufficiently large $n$. Then we obtain
\begin{equation}
\label{k3b}
\frac{1}{n p^2} \leq \frac{n}{(\log n + c_1)^2}
\end{equation}
and
\begin{equation}
\label{k3c}
u \leq  \frac{n^2 e^2}{2} \left(  \frac{\log n + c_2}{n} \right)^2 
{\rm exp}\left\{ - \frac{\log n + c_1}{2} \right\} = 
\frac{e^{2-c_1/2}}{2}  \frac{( \log n + c_2 )^2}{\sqrt{n}}.
\end{equation}
It follows from (\ref{k3a}), (\ref{k3b}) and (\ref{k3c}) that
\begin{eqnarray}
\varOmega & \leq & \frac{n}{(\log n + c_1)^2} \frac{e^{6 -3 c_1/2}}{8}  \frac{( \log n + c_2 )^6}{n \sqrt{n}} 
\left\{ 1 - \frac{e^{2-c_1/2}}{2}  \frac{( \log n + c_2 )^2}{\sqrt{n}} \right\}^{-1} \nonumber \\ 
& = & \frac{( \log n + c_2 )^6}{\sqrt{n} (\log n + c_1)^2} O(1) \rightarrow 0, \ \ \ n \rightarrow \infty.
\end{eqnarray}
\qed
\par
\medskip
\noindent
{\bf Proof of Proposition 4.2}
\par
\medskip
\noindent
As discussed above, for a bipartite graph $G_n = ({\cal V}_1,{\cal V}_2,{\cal E})$ with $\# V_1 = \# V_2 = n$, 
\begin{eqnarray}
& & \mathbb{P}(G_n \ {\rm has} \  {\rm at} \  {\rm least} \ {\rm  one} \ {\rm  isolated} \ 
{\rm  point}) \leq \mathbb{P}(G_n \ {\rm has} \ {\rm  no} \ {\rm  perfect} \ {\rm matchings}) \nonumber 
\\ 
& \leq & \mathbb{P}(G_n \ {\rm has} \  {\rm at} \  {\rm least} \ {\rm  one} \ {\rm  isolated} \ 
{\rm  point}) + \mathbb{P}\left(G_n \ {\rm  satisfies} \  {\bf Condition \ 4.11} \right) \nonumber \\ 
\end{eqnarray}
holds. We can see from {\bf Proposition 4.9}, {\bf Proposition 4.12} and the squeeze theorem that
\begin{eqnarray}
& & \lim_{n \rightarrow \infty} \mathbb{P}(G_n \ {\rm has} \ {\rm  no} \ {\rm  perfect} \ {\rm matchings}) 
\nonumber \\ 
& = & \lim_{n \rightarrow \infty} \mathbb{P}(G_n \ {\rm has} \  {\rm at} \  {\rm least} \ {\rm one} \ 
{\rm isolated} \ {\rm  point}) \nonumber \\ 
& = & 1 - e^{-\lambda}.
\end{eqnarray}
\qed

\subsection{Eigenvalue degeneracy probability for sparse random matrices}

We are now in a position to begin the proof of the main theorem. Suppose that 
random variables $Y_{j \ell}$ ($j,\ell=1,2,\cdots,N)$ independently obey Bernoulli 
distribution with a parameter $p$ and $K$-valued (broad) continuous random 
variables $Z_{j \ell}$ ($j,\ell = 1,2,\cdots,N$) are independently distributed. We want to 
evaluate the degeneracy probability for the eigenvalues of the $N \times N$ random matrix
$(Y_{j \ell} Z_{j \ell} )_{j,\ell=1,2,\cdots,N}$.
\par
\medskip
\noindent
{\bf Theorem 4.13}
\par
\medskip
\noindent
Let $c \in \mathbb{R}$. A parameter $p$ depends on $N$ and satisfies
\begin{equation}
p = p(N) = \frac{\log N + c}{N} + o\left( \frac{1}{N} \right), \ \ \ N \rightarrow \infty.
\end{equation}
Then 
\begin{equation}
\lim_{N \rightarrow \infty} 
\mathbb{P}\left({\rm the} \ {\rm matrix}  \ (Y_{jl} Z_{jl}) \ {\rm has} \  
N \ {\rm distinct} \  {\rm eigenvalues} \right) = 
e^{-\lambda} + \lambda e^{-\lambda}
\end{equation}
holds, where $\lambda = 2 e^{-c}$.
\par
\medskip
\noindent
{\bf Proof}
\par
\medskip
\noindent
We evaluate the probability for the corresponding 
random bipartite graph $G = ({\cal V}^R, {\cal V}^C, {\cal E})$ with 
${\cal E} = \{ (j,\ell) | Y_{j \ell} = 1 \}$ to satisfy {\bf Condition 4.1}, 
because it is equal to the probability that $N$ eigenvalues are all 
distinct. 
\par
Let us first bound the probability above. We write the number of 
isolated points of $G$ as $X_N$ and $\{ 1,2,\cdots,N \}$ as ${\cal I}_N$. 
Then we find
\begin{eqnarray}
& & \mathbb{P}(G \ {\rm satisfies} \ {\bf Condition \ 4.1}) 
\nonumber \\ 
& =  & 
\mathbb{P}(G \ {\rm has} \ {\rm a} \ {\rm  perfect} \ {\rm  matching}
\nonumber \\ 
& & {\rm or} \ \exists j \in {\cal I}_N, \ G[{\cal I}_N \setminus \{ j \}] \ {\rm has} 
\ {\rm a} \ {\rm perfect} \ 
{\rm matching} ).
\end{eqnarray}
As a necessary condition to have a perfect matching is the absence of isolated points, 
one can see that
\begin{eqnarray}
& & \mathbb{P}(G \ {\rm satisfies} \ {\bf Condition \ 4.1}) 
\nonumber \\ 
& \leq & 
\mathbb{P}(G \ {\rm has} \ {\rm no} \ {\rm  isolated} \ {\rm points} 
\nonumber \\ 
& & {\rm or} \ \exists j \in {\cal I}_N, G[{\cal I}_N \setminus \{ j \}] \ {\rm has} 
\ {\rm no} \ {\rm isolated} \ 
{\rm points} ) \nonumber \\ 
& =  & 
\mathbb{P}(X_N = 0) 
\nonumber \\ 
& & + 
\mathbb{P}(X_N \geq 1 \ {\rm and} \  \exists j \in {\cal I}_N, G[{\cal I}_N \setminus \{ j \}] \ {\rm has} \ {\rm no} \ {\rm isolated} \ 
{\rm points} ). \nonumber \\
\end{eqnarray}
When $X_N \geq 1$ and $\exists j \in {\cal I}_N$, $G[{\cal I}_N \setminus \{ j \}]$ has no isolated points,  
the isolated points of $G$ must be chosen from $\{v^R_j, v^C_j \}$. It follows that
\begin{eqnarray}
& & \mathbb{P}(G \ {\rm satisfies} \ {\bf Condition \ 4.1}) 
\nonumber \\ 
& \leq & \mathbb{P}(X_N \leq 1) + \mathbb{P}\left( \exists j \in {\cal I}_N, v^R_j \ 
{\rm and} \ v^C_j \ {\rm are} \ {\rm both} \ {\rm isolated} \ {\rm points} \right). \nonumber \\ 
\end{eqnarray}
Because of {\bf Proposition 4.9}, the first term of the RHS satisfies
\begin{equation}
\lim_{N \rightarrow \infty} \mathbb{P}(X_N \leq 1) = e^{-\lambda} + \lambda e^{-\lambda}.
\end{equation}
The second term of the RHS is estimated as
\begin{eqnarray}
& & \mathbb{P}\left( \exists j \in {\cal I}_N, v^R_j \ 
{\rm and} \ v^C_j \ {\rm are} \ {\rm both} \ {\rm isolated} \ {\rm points} \right)
\nonumber \\ 
& \leq & \sum_{j \in {\cal I}_N} \mathbb{P}\left(v^R_j \ {\rm and} \ v^C_j \ {\rm are} \
 {\rm both} \ {\rm isolated} \ {\rm points} \right) 
\nonumber \\ 
& = & \sum_{j \in {\cal I}_N} (1 - p)^{2 N - 1} = N (1 - p)^{2 N - 1}.
\end{eqnarray}
Then one can see that
\begin{eqnarray}
\label{2isolated}
& & \mathbb{P}\left( \exists j \in {\cal I}_N, v^R_j \ 
{\rm and} \ v^C_j \ {\rm are} \ {\rm both} \ {\rm isolated} \ {\rm points} \right)
\nonumber \\ 
& \leq & {\rm exp}\left\{ \log N  + (2 N - 1) \log(1 - p) \right\} \nonumber \\ 
& = & {\rm exp} \left[ \log N - (2 N - 1)  \left\{ \frac{\log N + c}{N} + o\left( 
\frac{1}{N} \right) \right\} \right] \nonumber \\ 
& = & {\rm exp}\left\{ - \log N + O(1) \right\} \rightarrow 0, \ \ \ N \rightarrow \infty.
\end{eqnarray}
Thus we obtain
\begin{equation}
\lim_{N \rightarrow \infty} \mathbb{P}(G \ {\rm satisfies} \ {\bf Condition \ 4.1})  
\leq  e^{-\lambda} + \lambda e^{-\lambda}.
\end{equation}
\par
Next we need to bound the probability below as
\begin{equation}
\lim_{N \rightarrow \infty} \mathbb{P}(G \ {\rm satisfies} \ {\bf Condition \ 4.1})  
\geq  e^{-\lambda} + \lambda e^{-\lambda},
\end{equation}
or equivalently
\begin{equation}
\label{notsatisfy}
\lim_{N \rightarrow \infty} \mathbb{P}(G \ {\rm does} \ {\rm not} \ {\rm satisfy} \ {\bf Condition \ 4.1})  
\leq  1 - e^{-\lambda} - \lambda e^{-\lambda}.
\end{equation}
In order to prove (\ref{notsatisfy}), we classify the cases according to the number of isolated points.
\par
\medskip
\noindent
(1) the case $X_N \geq 2$
\par
\medskip
\noindent
We can see from {\bf Proposition 4.9} that the corresponding asymptotic probability is
\begin{equation}
\lim_{N \rightarrow \infty} \mathbb{P}(X_N \geq 2) = 1 - e^{-\lambda} - \lambda e^{-\lambda}.
\end{equation}
Therefore we obtain
\begin{eqnarray}
\label{xn2}
& & \lim_{N \rightarrow \infty} \mathbb{P}(X_N \geq 2 \ {\rm and} \ G \ {\rm does} \ {\rm not} \ {\rm satisfy} \ {\bf Condition \ 4.1})  \nonumber \\ 
& \leq & \lim_{N \rightarrow \infty} \mathbb{P}(X_N \geq 2) = 1 - e^{-\lambda} - \lambda e^{-\lambda}.
\end{eqnarray}
\par
\medskip
\noindent
(2) the case $X_N = 1$
\par
\medskip
\noindent
Suppose that $v^R_k \in {\cal V}_R$ is the only isolated point of $G$. In order to make {\bf Condition 4.1} 
unsatisfied, a bipartite subgraph $G[{\cal I}_N \setminus \{k\}]$ must have no perfect matchings. 
Then $G[{\cal I}_N \setminus \{k\}]$ has at least an isolated point or {\bf Condition 4.11} holds. 
Let us separately consider those two cases.
\par
\medskip
(a) If $G[{\cal I}_N \setminus \{k\}]$ has an isolated point $w$, then $w$ is not an isolated point 
of $G$. Therefore $w \in {\cal V}_R$ and $(w,v^C_N) \in {\cal E}$.  It follows that 
the total probability to have such pairs $\{ v^R_k, w \}$ is equal to or less than
\begin{eqnarray}
& & \Pi_a(N) = \sum_{k=1}^N \mathbb{P}(v^R_k \ {\rm is} \ {\rm an} \ {\rm isolated} 
\ {\rm point} \ {\rm of} \ G)  
\nonumber \\ 
& & \times  \sum_{j \in {\cal I}_N \setminus \{k\}} 
\mathbb{P}((v^R_j,v^C_k) \in {\cal E}) 
\ \mathbb{P}(v^R_j \ {\rm is} \ {\rm an} \ {\rm isolated} 
\ {\rm point} \ {\rm of} \ G[{\cal I}_N \setminus \{k\}]). \nonumber \\
\end{eqnarray} 
We can see from
\begin{equation}
\mathbb{P}((v^R_j,v^C_k) \in {\cal E}) = p(N)
\end{equation}
and
\begin{equation}
\mathbb{P}(v^R_k \ {\rm is} \ {\rm an} \ {\rm isolated} 
\ {\rm point} \ {\rm of} \ G) = (1 - p)^N = \frac{\mathbb{E}[X_N]}{2 N}
\end{equation}
that
\begin{eqnarray}
\Pi_a(N)= \frac{1}{4} p(N) \mathbb{E}[X_N] \mathbb{E}[X_{N-1}].
\end{eqnarray} 
In the proof of {\bf Proposition 4.9}, we showed that  $\mathbb{E}[X_N] < \infty$. Moreover, as
\begin{equation}
p(N) = \frac{\log N + c}{N} + o\left( \frac{1}{N} \right) = 
p(N-1) + o\left(\frac{1}{N}\right),
\end{equation}
$\mathbb{E}[X_{N-1}] < \infty$ also holds. Therefore $\Pi_a(N)$ goes to 
zero in the limit $N \rightarrow \infty$.  We need to consider 
the other case when $v^C_k \in {\cal V}_C$ is the only isolated point of $G$. 
In that case, a similar argument can be applied and we find that the corresponding 
probability also goes to zero in the limit $N \rightarrow \infty$.
\par
\medskip
(b) Next we consider the case when  $v^R_k \in {\cal V}_R$ is the only isolated point 
of $G$ and $G[{\cal I}_N \setminus \{k\}]$ satisfies {\bf Condition 4.11}. The total probability 
of such cases is equal to or less than
\begin{eqnarray}
\Pi_b(N) & = & \sum_{k=1}^N \mathbb{P}(v^R_k \ {\rm is} \ {\rm an} \ {\rm isolated} 
\ {\rm point} \ {\rm of} \ G)  \nonumber \\ 
& & \times \mathbb{P}(G[{\cal I}_N \setminus \{k\}] \ {\rm satisfies} \ 
{\bf Condition \ 4.11}) \nonumber \\ 
& = & \frac{1}{2} \mathbb{E}[X_N] \ \mathbb{P}(G[{\cal I}_N \setminus \{k\}] \ {\rm satisfies} \ 
{\bf Condition \ 4.11}) .
\end{eqnarray}
We can again see from the proof of {\bf Proposition 4.9} that $\mathbb{E}[X_N] < \infty$. 
Since $p(N) = p(N-1) + o(1/N)$, {\bf Proposition 4.12} leads to 
\begin{equation}
\lim_{N \rightarrow \infty} \mathbb{P}(G[{\cal I}_N \setminus \{k\}] \ {\rm satisfies} \ 
{\bf Condition \ 4.11}) =0.
\end{equation}
Thus $\Pi_b(N)$ goes to zero in the limit $N \rightarrow \infty$.  The other case when 
$v^C_k \in {\cal V}_C$ can similarly be treated and we find that the corresponding 
probability also goes to zero in the limit $N \rightarrow \infty$.
\par
\medskip
\noindent
Putting the estimates in the cases (a) an (b) together, we find
\begin{eqnarray}
\label{xn1}
& & \lim_{N \rightarrow \infty} \mathbb{P}(X_N = 1 \ {\rm and} \ G \ {\rm does} \ {\rm not} \ {\rm satisfy} \ {\bf Condition \ 4.1})  \nonumber \\ 
& \leq & \lim_{N \rightarrow \infty} (2 \Pi_a(N) + 2 \Pi_b(N) )= 0.
\end{eqnarray}
\par
\medskip
\noindent
(3) the case $X_N = 0$
\par
\medskip
\noindent
If $X_N=0$, in order to make {\bf Condition 4.1} unsatisfied, {\bf Condition 4.11} must be satisfied.
It follows from {\bf Proposition 4.12} that
\begin{equation}
\lim_{N \rightarrow \infty} \mathbb{P}(G \ {\rm satisfies} \  {\bf Condition \ 4.11}) = 0.
\end{equation} 
Hence one can see that
\begin{eqnarray}
\label{xn0}
& & \lim_{N \rightarrow \infty} \mathbb{P}(X_N = 0 \ {\rm and} \ G \ {\rm does} \ {\rm not} \ {\rm satisfy} \ {\bf Condition \ 4.1})  \nonumber \\ 
& \leq & \lim_{N \rightarrow \infty} \mathbb{P}(G \ {\rm satisfies} \  {\bf Condition \ 4.11}) = 0.
\end{eqnarray}
\par
\medskip
\noindent
It follows from (\ref{xn2}), (\ref{xn1}) and (\ref{xn0}) that (\ref{notsatisfy}) holds. Therefore 
we arrive at
\begin{equation}
\lim_{N \rightarrow \infty} \mathbb{P}(G \ {\rm satisfies} \ {\bf Condition \ 4.1})  
=   e^{-\lambda} + \lambda e^{-\lambda}.
\end{equation}
This is the required result. \qed
\par
\medskip
\noindent
The dependence of the parameter $p$ on $N$ means that the random matrix is sparse.  
This theorem assures that the eigenvalue degeneracy probability can be positive 
in such a sparse limit.

\section{Eigenvalue degeneracy in symmetric sparse random matrices}
\setcounter{equation}{0}
\renewcommand{\theequation}{5.\arabic{equation}}

In this section, we consider a restricted case when the random matrices are 
symmetric. Let us put $K=\mathbb{R}$ or $K = \mathbb{C}$. We suppose that 
$Z_{j \ell}$ ($1 \leq j \leq \ell \leq N$) 
are independent $K$-valued (broad) continuous random variables. Moreover we define $Y_{j \ell}$ ($1 \leq j \leq \ell \leq N$) 
as independent random variables obeying Bernoulli distribution with
\begin{equation}
\mathbb{P}(Y_{j \ell} = 1) = \left\{ \begin{array}{ll} p, & {\rm if} \ j < \ell, \\
q, & {\rm if} \ j = \ell. \end{array} \right.
\end{equation}
For $1 \leq \ell < j \leq N$, we assume $Z_{j \ell} = Z_{\ell j}$ and $Y_{j \ell} = Y_{\ell j}$. 
Using these random variables, we consider the probability for a symmetric random matrix 
$(Y_{j \ell} Z_{j \ell})_{j,\ell=1,2,\cdots,N}$ to have $N$ distinct eigenvalues. As is well-known, 
in the case $K = \mathbb{R}$,  all the eigenvalues are real.
\par
To begin with, we arbitrarily fix $y_{j \ell} \in \{0,1\}$ ($1 \leq j \leq \ell \leq N$)
and study the conditional probability under the condition $Y_{j \ell}= y_{j \ell}$. 
It follows from {\bf Theorem 2.5} that this conditional probability is zero or one, 
and the necessary and sufficient condition to have the conditional probability one is:
\par
\medskip
\noindent
There exists at least one set $\{ z_{j \ell} \}_{1 \leq j \leq \ell \leq N}$ with $z_{j \ell} \in K$, 
so that the matrix $(y_{j \ell} z_{j \ell})_{j \ell=1,2,\cdots,N}$  has $N$ distinct eigenvalues. 
Here we assume $z_{j \ell} = z_{\ell j}$ and $y_{j \ell} = y_{\ell j}$ for $1 \leq \ell < j \leq N$.
\par
\medskip
\noindent
Therefore, as before, the probability for the matrix $(Y_{j \ell} Z_{j \ell})_{j,\ell=1,2,\cdots,N}$ 
to have $N$ distinct eigenvalues does not depend on the distributions of $Z_{j \ell}$ and 
depends only on $N$, $p$ and $q$. Moreover,  from {\bf Theorem 3.11}, one can see that 
such a probability is equal to the probability for the random bipartite graph 
$G = ({\cal V}^R, {\cal V}^C, \mathfrak{E})$, $\mathfrak{E} = \{ \langle j,\ell \rangle | Y_{j \ell} = 1 \}$ 
to satisfy {\bf Condition 4.1}, when ${\cal E} = \mathfrak{E}$.
\par
Let us now find the number distribution of the isolated points. Since the edges in $\mathfrak{E}$ 
satisfy
\begin{equation}
\langle j,\ell \rangle \in \mathfrak{E} \Longleftrightarrow \langle \ell,j \rangle \in \mathfrak{E},
\end{equation}
we obtain
\begin{equation}
v_j^R \ {\rm is} \ {\rm an} \ {\rm isolated} \ {\rm  point} \Longleftrightarrow 
v_j^C \ {\rm is} \ {\rm an} \ {\rm isolated} \ {\rm  point}.
\end{equation}
Therefore, using the maps
\begin{equation}
j \in {\cal I}_N \mapsto v^R_j \in {\cal V}^R \ {\rm and} \ 
j \in {\cal I}_N \mapsto v^C_j \in {\cal V}^C,
\end{equation}
we identify ${\cal V}^R$, ${\cal V}^C$ and ${\cal I}_N = \{ 1,2,\cdots,N \}$. Accordingly 
we say $j \in {\cal I}_N$ is an isolated point, when $v^R_j$ and $v^C_j$ are 
isolated points. Defining a random variable $X$ as
\begin{equation}
X = \#\{ j \in {\cal I}_N | \ j \ {\rm is} \ {\rm an} \ {\rm isolated} \ {\rm point} \},
\end{equation}
we find the following lemma.
\par
\medskip
\noindent
{\bf Lemma 5.1}
\par
\medskip
\noindent
Let $c \in \mathbb{R}$ be a constant. The probability parameters $p$ and $q$ depend on $N$ as
\begin{equation}
p = p(N) = \frac{\log N + c}{N} + o\left( \frac{1}{N} \right), \ \ \ N \rightarrow \infty, 
\end{equation} 
\begin{equation}
\lim_{N \rightarrow \infty} q(N) = q_\infty \in [0,1].
\end{equation}
Then the probability to have $X =x$ is asymptotically evaluated as
\begin{equation}
\lim_{n \rightarrow \infty} \mathbb{P}(X = x) = e^{-\mu} \frac{\mu^x}{x!}
\end{equation}
with $\mu = (1 - q_\infty) e^{-c}$.
\par
\medskip
\noindent
{\bf Proof}
\par
\medskip
\noindent
Let us define random variables 
\begin{equation}
B_j = \left\{ \begin{array}{ll} 1, & {\rm if} \ j \ {\rm is} \ {\rm an} \ {\rm isolated} \ {\rm  point},  \\ 
0, & {\rm otherwise} \end{array} \right.
\end{equation}
($j = 1,2,\cdots,N$) and choose a subset $I \subset {\cal I}_N$ with $\# I =  k \in \mathbb{N}$. Then
\begin{eqnarray}
\mathbb{E}\left( \prod_{j \in I} B_j \right) & = & \mathbb{P}( j \in I \ {\rm are} \  {\rm all} 
\ {\rm  isolated} \ {\rm points}) \nonumber \\
& = & \prod_{h=1}^k \left\{ (1 - q) (1 - p)^{k-h} (1 - p)^{N - k} \right\} \nonumber \\ 
& = & (1 - q)^k (1 - p)^{k(k-1)/2} (1 - p)^{k(N-k)}.
\end{eqnarray}
It follows from {\bf Proposition 4.7} that
\begin{eqnarray}
\mathbb{E}\binom{X}{k} & = & \sum_{\substack{I \subset {\cal I}_N \\ \# I = k}} \mathbb{E} \left(
\prod_{j \in I} B_j \right) \nonumber \\ 
& = & \binom{N}{k} (1 - q)^k (1 - p)^{k(k-1)/2} (1 - p)^{k(N-k)}.
\end{eqnarray}
In the limit $N \rightarrow \infty$ with $k \in \mathbb{N}$ fixed, we find
\begin{eqnarray}
\mathbb{E}\binom{X}{k} & = & (1 - q)^k \frac{N^k \left\{ 1 +  o(1) \right\}}{k!}  {\rm exp}\left[ \left\{  k N + O(1) \right\} \log(1 - p)  \right] \nonumber \\ 
& = & (1 - q)^k \frac{N^k \left\{ 1 +  o(1) \right\}}{k!} {\rm exp}\left\{ - k \log N - c k  + o(1) \right\} 
\nonumber \\ 
& = & \frac{(1 - q_\infty)^k}{k!} {\rm exp}( - c k)  + o(1) = \frac{\mu^k}{k!} + o(1), 
\end{eqnarray}
which also holds for $k =0$. Thus one can see from {\bf Proposition 4.6} that
\begin{eqnarray}
\lim_{N \rightarrow \infty} \mathbb{P}(X = x) & = & \sum_{j=x}^\infty (-1)^{j-x} \binom{j}{x} \frac{\mu^j}{j!} 
= \sum_{j=x}^\infty (-1)^{j-x} \frac{\mu^j}{x! (j-x)!} \nonumber \\ 
& = & \frac{\mu^x}{x!} \sum_{\ell=0}^\infty \frac{(-\mu)^\ell}{\ell!} = \frac{\mu^x}{x!} e^{-\mu}.
\end{eqnarray}
\qed
\par
\medskip
\noindent
In the case $q_\infty = 0$, {\bf Lemma 5.1} is reduced to a known result\cite{FK} (Theorem 3.1).
As before, we now need to consider the consequence of the absence of perfect matchings. 
For that purpose, we first show the following lemma.
\par
\medskip
\noindent
{\bf Lemma 5.2}
\par
\medskip
\noindent
Let us suppose that a bipartite graph $G = ({\cal V}_1, {\cal V}_2, \mathfrak{E})$ with $\#{\cal V}_1 = 
\#{\cal V}_2 = n$ satisfies
\begin{equation}
\langle j,\ell \rangle \in \mathfrak{E} \Longleftrightarrow \langle \ell,j \rangle \in \mathfrak{E}
\end{equation}
and $G$ has no perfect matchings. We introduce a notation
\begin{equation}
{\tilde \Gamma}_G(A) = \{ \ell \in {\cal I}_N | \exists j \in A, \langle j,\ell \rangle \in \mathfrak{E} \}
\end{equation}
for $A \subset {\cal I}_N$. As before, if $A$ is a one-point set $\{ j \}$, then 
${\tilde \Gamma}_G(\{ j \})$ is simply written as ${\tilde \Gamma}_G(j)$.
Among the sets $A \subset {\cal I}_N$ satisfying $\# {\tilde \Gamma}_G(A) \leq \# A - 1$, we choose one set minimizing $\# A$ and call it $I$. Then $I$ satisfies the followings.
\par
\medskip
\noindent
(1) $\# {\tilde \Gamma}_G(I) = \# I - 1$.
\par
\medskip
\noindent
(2) $ \# I \leq (n + 1)/2$.
\par
\medskip
\noindent
(3) $\# ({\tilde \Gamma}_G(\ell) \cap I) \geq 2$ for arbitrary $\ell \in {\tilde \Gamma}_G(I)$. 
\par
\medskip
\noindent
(4) $I \cap {\tilde \Gamma}_G(I) = \emptyset$.
\par
\medskip
\noindent
{\bf Proof}
\par
\medskip
\noindent
We can prove (1), (2) and (3) in the same way as {\bf Proposition 4.10}. A proof of (4) is given below. 
\par
Let us define $S = I \cap {\tilde \Gamma}_G(I)$ and $J = I \setminus S$. When we arbitrarily take 
$j \in J$ and $s \in S$, we find $j \notin {\tilde \Gamma}_G(I)$ and $s \in I$. Therefore 
$\langle s, j \rangle \notin \mathfrak{E}$, so that $\langle j,s \rangle \notin \mathfrak{E}$ or 
${\tilde \Gamma}_G(J) \cap S = \emptyset$.
\par
On the other hand, it follows from $J \subset I$ that ${\tilde \Gamma}_G(J) \subset {\tilde \Gamma}_G(I)$. 
Using this inclusion relation with ${\tilde \Gamma}_G(J) \cap S = \emptyset$, we find ${\tilde \Gamma}_G(J) \subset 
{\tilde \Gamma}_G(I)  \setminus S$. Noting $S = I \cap {\tilde \Gamma}_G(I)$ and (1), we obtain
\begin{eqnarray}
\# {\tilde \Gamma}_G(J) & \leq & \# ({\tilde \Gamma}_G(I) \setminus S) = \# {\tilde \Gamma}_G(I) - \# S 
\nonumber \\ 
& = & \# I - 1 - \# S = \#(I \setminus S) - 1 = \# J - 1.
\end{eqnarray}
Among the sets $A \subset {\cal I}_N$ satisfying $\# {\tilde \Gamma}_G(A) \leq \# A - 1$, $I$ is 
chosen as the set minimizing $\# A$.  Therefore we need to have $\# I \leq \# J$. 
Then we find $I = J$ or $S = \emptyset$, because $J \subset I$. \qed
\par
\medskip
\noindent
As a consequence of {\bf Lemma 5.2}, if a bipartite graph $G$ has no perfect matchings, it has at least one isolated point or satisfies the following condition. 
\par
\medskip
\noindent
{\bf Condition 5.3}
\par
\medskip
\noindent
There exists $k \in \mathbb{N}$ ($2 \leq k \leq (n+1)/2$) satisfying 
$I \subset {\cal I}_n$, $J \subset {\cal I}_n \setminus I$ with $\# I = k$, $\# J = k-1$, 
so that ${\tilde \Gamma}_G(I) = J$ and $\forall \ell \in J$, $\# ({\tilde \Gamma}_G(\ell) \cap I )\geq 2$.
\par
\medskip
\noindent
The case $k=1$ again corresponds to graphs with at least one isolated point. 
The probability for {\bf Condition 5.3} to hold is asymptotically evaluated as follows.
\par
\medskip
\noindent
{\bf Lemma 5.4}
\par
\medskip
\noindent
Suppose that $\# {\cal V}_1 = \# {\cal V}_2 = n$ and the maps 
\begin{equation}
v_1: {\cal I}_n \mapsto {\cal V}_1 \ {\rm and} \ v_2:{\cal I}_n \mapsto {\cal V}_2
\end{equation}
are bijections. Each $(j,\ell)$ ($1 \leq j < \ell \leq n$) independently satisfies 
$(v_1(j),v_2(\ell)) \in \mathfrak{E} \subset {\cal V}_1 \times {\cal V}_2$ with probability $p$. 
For $(j,\ell)$ ($1 \leq \ell < j \leq n$), we set $(v_1(j),v_2(\ell)) \in \mathfrak{E} 
\Longleftrightarrow (v_1(\ell),v_2(j)) \in \mathfrak{E}$. The 
probability parameter $p$ depends on $n$ as
\begin{equation}
p = p(n) = \frac{\log n + c}{n} + o\left( \frac{1}{n} \right), \ \ \ n \rightarrow \infty
\end{equation} 
with a constant $c \in \mathbb{R}$. Then the random bipartite graph 
$G_n = ({\cal V}_1,{\cal V}_2, \mathfrak{E})$ fulfills
\begin{equation}
\lim_{n  \rightarrow \infty} \mathbb{P}\left(G_n \ {\rm  satisfies} \  {\bf Condition \ 5.3} \right) = 0.
\end{equation}
\par
\medskip
\noindent
{\bf Remark}
\par
\medskip
\noindent
In {\bf Lemma 5.4}, the probabilities to have $(v_1(j),v_2(j)) \in \mathfrak{E}$ for $1 \leq j \leq n$ are arbitrary.
Even independence is not required for them.
\par
\medskip
\noindent
{\bf Proof of Lemma 5.4}
\par
\medskip
\noindent
Let us first fix $k \in \mathbb{N}$ satisfying $2 \leq k \leq (n + 1)/2$. 
We choose vertex sets $I \subset {\cal I}_n$ and $J \subset {\cal I}_n \setminus I$ with $\# I = k$ and $\# J = k - 1$. There are $\displaystyle \binom{n}{k} \binom{n-k}{k-1}$ ways to choose such $\{ I, J\}$. Then the probability to have $\langle j,\ell \rangle \notin \mathfrak{E}$ for all 
$(j,\ell) \in I \times \left( {\cal I}_n \setminus J \right)$ is  
\begin{equation}
\prod_{h=1}^k (1 - p)^{k-h} (1 - p)^{n - 2 k +1}= (1 - p)^{k (k - 1)/2 + k( n - 2 k + 1)},
\end{equation}
so that
\begin{equation}
\mathbb{P}({\tilde \Gamma}_G(I) = J) \leq (1 - p)^{k (k - 1)/2 + k( n - 2 k + 1)}
\end{equation}
for each $\{ I,J \}$. 
\par
Next we fix $\ell \in J$ and choose $j,j' \in I$ ($j \neq j'$). For each $\ell$ there are $\displaystyle \binom{k}{2}$ ways to choose such $j,j'$. The probability to have $\langle j,\ell \rangle \in 
\mathfrak{E}$ and 
$\langle j',\ell \rangle \in \mathfrak{E}$ for each triple $\{ \ell,j,j' \}$ is $\displaystyle p^2$. Thus we obtain
\begin{equation}
\mathbb{P}\left( \exists \{j,j'\}, \langle j,\ell \rangle \in \mathfrak{E} \ {\rm and} \ 
\langle j',\ell \rangle\in \mathfrak{E} \right) \leq \binom{k}{2} p^2
\end{equation}
for each $\ell$.
\par
It can be seen from the above argument that
\begin{eqnarray}
& & \mathbb{P}\left(G_n \ {\rm  satisfies} \  {\bf Condition \ 5.3} \right) \nonumber \\
& \leq & 
\sum_{k=2}^{\lfloor (n+1)/2 \rfloor} \binom{n}{k} \binom{n-k}{k-1} \left\{ \binom{k}{2} p^2 \right\}^{k-1} 
(1 - p)^{k( k - 1)/2 + k(n - 2 k + 1)} \nonumber \\ 
& \leq & \sum_{k=2}^{\lfloor (n+1)/2 \rfloor} \left( \frac{n e}{k} \right)^k
\left( \frac{n e}{k - 1} \right)^{k-1}  \left\{ \binom{k}{2} p^2 \right\}^{k-1} 
{\rm exp}\left\{ -p k \left(n - \frac{3}{2} k + \frac{1}{2} \right) \right\} \nonumber \\
& = & \frac{2}{n e p^2} \sum_{k=2}^{\lfloor (n+1)/2 \rfloor} \frac{1}{k} \left( \frac{(n e p)^2}{2} \right)^k 
{\rm exp}\left\{ - p k \left( n - \frac{3}{2} k + \frac{1}{2} \right) \right\}.
\end{eqnarray}
When $2 \leq k \leq 4$, one obtains
\begin{eqnarray}
& & \frac{2}{n e p^2} \frac{1}{k} \left( \frac{(n e p)^2}{2} \right)^k 
{\rm exp}\left\{ - p k \left( n - \frac{3}{2} k + \frac{1}{2} \right) \right\} \nonumber \\ 
& = & O\left( \frac{\displaystyle ( \log n )^{2(k-1)}}{\displaystyle n^{k-1}} \right) \rightarrow 0, \ \ \ n \rightarrow \infty.
\end{eqnarray}
As we can see from $k \leq (n + 1)/2$ that
\begin{equation}
{\rm exp}\left\{ - p \left( n - \frac{3}{2} k + \frac{1}{2} \right) \right\} \leq 
{\rm exp}\left\{ - \frac{p}{4} (n - 1) \right\},
\end{equation}
the sum of the terms with $k \geq 5$ satisfies
\begin{eqnarray}
\label{k5a}
\varOmega & =  & 
\frac{2}{n e p^2} \sum_{k=5}^{\lfloor (n+1)/2 \rfloor} \frac{1}{k} \left( \frac{(n e p)^2}{2} \right)^k 
{\rm exp}\left\{ - p k \left( n - \frac{3}{2} k + \frac{1}{2} \right) \right\} \nonumber \\ 
& \leq & \frac{1}{2 n p^2}  \sum_{k=5}^{\lfloor (n+1)/2 \rfloor}  u^k \leq \frac{1}{2 n p^2} \frac{u^5}{1 - u},
\end{eqnarray}
where
\begin{equation}
u = \frac{(n e p)^2}{2} e^{-p (n-1) /4}.
\end{equation}
Now we take constants $c_1,c_2 \in \mathbb{R}$ complying with
\begin{equation}
\frac{\log n + c_1}{n} \leq p(n) \leq \frac{\log n + c_2}{n}
\end{equation}
for sufficiently large $n$. It follows that
\begin{equation}
\label{k5b}
\frac{1}{n p^2} \leq \frac{n}{(\log n + c_1)^2}
\end{equation}
and
\begin{equation}
\label{k5c}
u \leq  \frac{n^2 e^2}{2} \left(  \frac{\log n + c_2}{n} \right)^2 
{\rm exp}\left\{ - \frac{\log n + c_1}{4} + \frac{p}{4}  \right\} = 
\frac{e^{2+(p-c_1)/4}}{2}  \frac{( \log n + c_2 )^2}{\sqrt[4]{n}}.
\end{equation}
We can see from (\ref{k5a}), (\ref{k5b}) and (\ref{k5c}) that
\begin{eqnarray}
\varOmega & \leq & \frac{n}{2 (\log n + c_1)^2} 
\frac{e^{10+5 (p-c_1)/4}}{32}  \frac{( \log n + c_2 )^{10}}{n \sqrt[4]{n}}
\left\{ 1 - \frac{e^{2+(p-c_1)/4}}{2}  \frac{( \log n + c_2 )^2}{\sqrt[4]{n}} \right\}^{-1} 
\nonumber \\ 
& = & \frac{( \log n + c_2 )^{10}}{\sqrt[4]{n} (\log n + c_1)^2} O(1) \rightarrow 0, \ \ \ n \rightarrow \infty.
\end{eqnarray}
\qed
\par
We are now in a position to asymptotically evaluate the probability for a symmetric random matrix 
$(Y_{j \ell} Z_{j \ell})_{j,\ell=1,2,\cdots,N}$ to have $N$ distinct eigenvalues. 
\par
\medskip
\noindent
{\bf Theorem 5.5}
\par
\medskip
\noindent
Let $c \in \mathbb{R}$. Probability parameters $p,q$ depend on $N$ and satisfy
\begin{equation}
p = p(N) = \frac{\log N + c}{N} + o\left( \frac{1}{N} \right), \ \ \ N \rightarrow \infty.
\end{equation}
\begin{equation}
\lim_{N \rightarrow \infty} q(N) = q_\infty \in [0,1].
\end{equation}
Then we obtain
\begin{equation}
\lim_{N \rightarrow \infty} 
\mathbb{P}\left({\rm the} \ {\rm matrix}  \ (Y_{j \ell} Z_{j \ell}) \ {\rm has} \  
N \ {\rm distinct} \  {\rm eigenvalues} \right) = 
e^{-\mu} + \mu e^{-\mu},
\end{equation}
where $\mu = (1 - q_\infty)  e^{-c}$.
\par
\medskip
\noindent
{\bf Proof}
\par
\medskip
\noindent
It is sufficient to give the probability for the random bipartite graph 
$G = ({\cal V}^R, {\cal V}^C, \mathfrak{E})$, $\mathfrak{E} = \{ \langle j,\ell \rangle | Y_{j \ell} = 1 \}$
to satisfy {\bf Condition 4.1}, when ${\cal E} = \mathfrak{E}$. Let us put 
$X = \# \{ j \in {\cal I}_N | j \ {\rm is} \ {\rm an} \ {\rm isolated} \ {\rm point} \}$ 
and classify the cases according to the value of $X$.
\par
\medskip
\noindent
(1) When $X \geq 2$, {\bf Condition 4.1} is not satisfied.
\par
\medskip
\noindent
(2) When $X = 0$, if {\bf Condition 5.3} is not satisfied, then 
{\bf Condition 4.1} is satisfied. It follows from {\bf Lemma 5.4} that 
\begin{equation}
\lim_{N \rightarrow \infty} \mathbb{P}( {\bf Condition \ 5.3} \ {\rm is} \ 
{\rm satisfied} ) = 0.
\end{equation}
This result and {\bf Lemma 5.1} lead to
\begin{equation}
\lim_{N \rightarrow \infty} \mathbb{P}( X=0 \ {\rm and} \ {\bf Condition \ 4.1} \ {\rm is} \ 
{\rm satisfied} ) = e^{-\mu}.
\end{equation}
\par
\medskip
\noindent
(3) When $X = 1$, suppose that $j  \in {\cal I}_N$ is the isolated point. 
If the partial bipartite graph $G[{\cal I}_N \setminus \{j\}]$ has a perfect matching, then 
{\bf Condition 4.1} is satisfied. Let us consider each of  the following two cases when 
$G[{\cal I}_N \setminus \{j\}]$ has no perfect matchings. 
\par
\medskip
(a) First we consider the case when $G[{\cal I}_N \setminus \{j\}]$ has an isolated point 
$\ell \in {\cal I}_N \setminus \{j\}$, Since $\ell$ is not an isolated point of $G$, we obtain 
$\langle j,\ell \rangle \in \mathfrak{E}$. 
This is however contradictory to the fact that $j$ is an isolated point of $G$. Therefore 
it is impossible for $G[{\cal I}_N \setminus \{j\}]$ to have an isolated point.
\par
\medskip
(b) Next we consider the case when $G[{\cal I}_N \setminus \{j\}]$ satisfies {\bf Condition 5.3}. 
Since $j \in {\cal I}_N$ is an isolated point of $G$, $G$ itself also satisfies 
{\bf Condition 5.3}. One can see from {\bf Lemma 5.4} that the  probability for $G$ to satisfy 
{\bf Condition 5.3} converges to zero in the limit $N \rightarrow \infty$.
\par
\medskip
\noindent
The arguments on the cases (a) and (b) together with {\bf Lemma 5.1} yield
\begin{equation}
\lim_{N \rightarrow \infty} \mathbb{P}( X=1 \ {\rm and} \ {\bf Condition \ 4.1} \ {\rm is} \ 
{\rm satisfied} ) = \mu e^{-\mu}.
\end{equation}
Then we can conclude from the consideration of all the cases (1), (2) and (3) that
\begin{equation}
\lim_{N \rightarrow \infty} \mathbb{P}({\bf Condition \ 4.1} \ {\rm is} \ 
{\rm satisfied} ) = e^{-\mu} + \mu e^{-\mu}.
\end{equation}
This is equal to the asymptotic probability for a symmetric random matrix 
$(Y_{j \ell} Z_{j \ell})_{j,\ell=1,2,\cdots,N}$ to have $N$ distinct eigenvalues. \qed

\section{Discussion}

In this paper, the eigenvalue degeneracy probability is treated, when the probability measure
of the random matrix entries is continuous except at the origin. As a result, we obtain 
a positive degeneracy probability for the zero eigenvalue of a sparse random matrix model. On 
the other hand, when the probability measure has discontinuous points other than the origin, 
similar results are not known. This limitation should be overcome by analyzing 
more general discontinuous models.

\section*{Acknowledgement}
The author sincerely thanks Taro Nagao for valuable discussions, comments, and generous support, which greatly contributed to the clarity and presentation of this paper.

\end{document}